\documentclass[journal]{IEEEtran}

\usepackage{algorithmic}            
\usepackage{algorithm}

\usepackage[space]{cite}

\ifCLASSINFOpdf
   \usepackage[pdftex]{graphicx}
\else
 \usepackage[dvips]{graphicx}
\fi
\usepackage{graphicx}

\usepackage[cmex10]{amsmath}
\usepackage{txfonts}
\usepackage[mathscr]{euscript}
\usepackage{amsfonts}
\usepackage{bm}
\usepackage{longtable}

\usepackage{setspace}
\usepackage{rotating}
\usepackage{amsfonts}
\usepackage{tabularx}
\usepackage{enumerate}
\usepackage[table]{xcolor}
 \usepackage{epsfig}
\usepackage{epstopdf}
\usepackage[T1]{fontenc}
\usepackage{mwe}    
\usepackage{xcolor,etoolbox}
\usepackage{color,soul}
\usepackage{yfonts}
\usepackage{xr}
\usepackage{url}
\usepackage[normalem]{ulem}
\usepackage{subfig}
\usepackage{rotating}
\usepackage{pdflscape}

\newcommand\numberthis{\addtocounter{equation}{1}\tag{\theequation}}

\makeatletter 
\pretocmd\@bibitem{\color{black}\csname keycolor#1\endcsname}{}{\fail}
\newcommand\citecolor[1]{\@namedef{keycolor#1}{\color{black}}}
\makeatother
\citecolor{souto2016mimo}
\citecolor{wen2020time}
\citecolor{wen2019optimization}
\citecolor{jana2018dual}
\citecolor{jana2017pre}
\citecolor{li2020code}
\citecolor{li2017reduced}
\citecolor{rusek2011optimal}

\usepackage{epstopdf}					
\begin{document}
\title{A Novel Low Complexity Faster-than-Nyquist (FTN) Signaling Detector for Ultra High-Order QAM}
\author{Ahmed~Ibrahim, Ebrahim Bedeer, and Halim Yanikomeroglu
	
\thanks{A patent application incorporating parts of this paper has been filed \cite{ourPat}.}
}
	
\maketitle
\begin{abstract}
Faster-than-Nyquist (FTN) signaling is a promising non-orthogonal pulse modulation technique that can improve the spectral efficiency (SE) of next generation communication systems at the expense of higher detection complexity  to remove the introduced inter-symbol interference (ISI). In this paper, we investigate the detection problem of ultra high-order quadrature-amplitude modulation (QAM) FTN signaling where we exploit a mathematical programming technique based on the alternating directions multiplier method (ADMM).  The proposed ADMM sequence estimation (ADMMSE) FTN signaling detector demonstrates an excellent trade-off between performance and computational effort enabling successful detection and SE gains for QAM modulation orders as high as 64K (65,536). The complexity of the proposed ADMMSE detector is polynomial in the length of the transmit symbols sequence and its sensitivity to the modulation order increases only logarithmically. Simulation results show that for 16-QAM, the proposed ADMMSE FTN signaling detector achieves comparable SE gains to the generalized approach semidefinite relaxation-based sequence estimation (GASDRSE) FTN signaling detector, but at an experimentally evaluated much lower computational time. Simulation results additionally show SE gains for modulation orders starting from 4-QAM, or quadrature phase shift keying (QPSK), up to and including  64K-QAM when compared to  conventional Nyquist signaling. The very low computational effort required makes the proposed ADMMSE detector a practically promising FTN signaling detector for both low order and ultra high-order QAM FTN signaling systems. 

\end{abstract}


\section{Introduction}
Next generation wireless communication systems are expected to support novel use cases that are mainly driven by the ongoing and rapid changes in our societies and their impact in our lifestyle. While such changes are due to many contributing reasons, currently we witness the widespread of the COVID-19 pandemic that has certainly affected many aspects of our life and we relied more than ever on technology and video conferencing to support economy, online education, services, etc., which increases the demands on higher throughput. While the ongoing deployment of 5G wireless communication systems can support enhanced multimedia applications of peak rates of 20 Gbps \cite{itu2017minimum}, it is expected that such applications will evolve to augmented reality, 3DTV/holographic communications, multi-sense communications, and/or their combinations. The evolved applications are expected to require peak rates in the order of a few terabits per second (Tbps) which is beyond the capability of 5G systems \cite{itu2020, 8970173}. To support such requirements, we generally need larger bandwidth (up to 1 THz) than what is currently offered in 5G; however, such new frequency bands suffer from very high absorption \cite{8387210} and may be suitable only for short (few meters) range communication applications. Even the spectrum windows with low absorption, and hence, better propagation conditions, will not be fully available for next generation communications as they are already allocated by ITU to other services such as astronomy and earth exploration communications \cite{8387211}. That said, some wireless applications will still be offered over the existing 4G/5G frequency bands and there is a need for innovative solutions to improve the spectral efficiency (SE) given the limited bandwidth. 

Improving the SE is arguably the oldest problem in PHY from 1920s (predating Shannon) and gains of fractions of dB in signal-to-noise ratio (SNR) to maintain the same SE are of high value.  Faster-than-Nyquist (FTN) signaling is a promising technique for increasing SE in next generation communication systems compared to the classical Nyquist signaling \cite{anderson2013faster}. Nyquist showed that signaling at rates greater than $\frac{1}{T}$ of $T$-orthogonal pulses, i.e., pulses that are orthogonal to an $nT$ shift of themselves for nonzero integer $n$, results in ISI at the samples of receiver’s matched filter output \cite{nyquist1928certain}. 
On the other hand, FTN signaling accelerates the pulses beyond the Nyquist limit, which introduces inter-symbol interference (ISI).
The prominent work of J. E. Mazo in 1975 \cite{mazo1975faster} was the first to show that FTN signaling does not reduce the normalized minimum Euclidean distance when accelerating sinc pulses at a rate of  $\frac{1}{\tau }$, $ \tau \in \left[0.802, 1\right]$, higher than the Nyquist signaling, which was later known in the literature as Mazo limit.
The potential of FTN signaling extends beyond the Mazo limit in practical coded communications systems, where it was shown that substantial acceleration of the transmited pulses increases the constrained capacity, if the complexity bottleneck can be sorted out \cite{anderson2013faster}. This translates to profound returns at the cost of some mild performance loss. 

There are a number of communications technologies and standards that adopt ultra high-order quadrature-amplitude modulation (QAM) and for which additional SE gains can be achieved by using FTN signaling. These systems include high speed point-to-point microwave links such as RAy3 (the third generation of RAy) and digital video broadcasting DVB-C2 that use QAM modulation orders up to 4096 \cite{RACOM,5523733}, as well as the broadband cable based internet DOCSIS 3.1 standard that uses QAM modulation orders up to 16,384 (16K)\cite{8808157}. 
{  Using existing FTN signaling sequence estimation techniques to detect ultra high-order QAM FTN signaling will results in prohibitive computational complexity for practical implementations. For instance, increasing the modulation order for up to 64K (65536)-QAM will significantly increase the tree size in tree search based techniques or will significantly increase the trellis size in trellis search based techniques. That said, it is crucial to investigate novel sequence estimation techniques where their operation and complexity are insensitive to the modulation order. 
{It is worth noting that ISI elimination can either be achieved using equalizers/detectors at the receiver only, or using precoding techniques at the transmitter, or both.  For example, a variant of the G-to-minus-half (GTMH) precoding in \cite{8954896} that combines cyclic prefix, cyclic suffix, and discrete Fourier transform, can achieve a Nyquist-like bit-error rate (BER) performance for modulation orders up to 128/256 amplitude phase shift keying (APSK) at low complexity. In our work, we aim to detect ultra high-order QAM beyond the capabilities of the GTMH precoding, and we propose an efficient sequence estimation technique at the receiver.}
{It is worth noting that a number of works in the literature made progress towards the detection of high-order modulation FTN signaling \cite{8954896, wen2021waveform, song2020receiver}. For instance, the work in \cite{8954896} reported promising results of using a variant of the G-to-minus-half (GTMH) precoding to detect 128/256 amplitude phase shift keying (APSK) at low complexity. In \cite{wen2021waveform}, the authors showed that in the presence of phase noise precoded 4K-QAM FTN signaling has 5.8 dB SNR gain when compared to 16K-QAM Nyquist signaling.}

In this paper, we propose a low complexity polynomial time detection scheme that stems from the operations research and is known as the \textit{alternating directions multiplier method} (ADMM).} {\color{black}It is worthy to mention that ADMM has been used in the detection of MIMO transmission in time dispersive channels in \cite{souto2016mimo}}. Our  proposed ADMM detection algorithm demonstrates a very favorable combination of computational efficiency and performance which practically enables the detection of ultra high-order QAM modulation. The first polynomial complexity high-order QAM FTN signaling detector is the generalized approach semidefinite relaxation-based sequence estimation (GASDRSE) that was proposed in \cite{bedeer2017low} based on the semi-definite relaxation and Gaussian randomization. The GASDRSE algorithm showed excellent performance and SE gains for modulation orders up to 16-QAM. {However for modulation orders larger than 16, the algorithm did not provide satisfactory SE gains, besides requiring an impractically large computational time.} {  The GASDRSE will fail to provide satisfactory performance for ultra high-order QAM FTN signaling as it represents the constellation alphabets as a high degree polynomial constraint. This will significantly increase the constraints' matrices sizes in the case of ultra high-order QAM, and eventually the SDR solver fails to find a solution.} Our simulation results show that the proposed ADMMSE detector in this paper can outperform GASDRSE for 16-QAM and quadrature phase-shift keying (QPSK) while requiring less than 25$\%$ of the computational time. Moreover, the SE gains of the proposed ADMMSE detector are up to 44.7\% higher than the GASDRSE detector for 16-QAM FTN signaling. More importantly, the proposed ADMMSE FTN signaling detector in this paper succeeds in achieving SE gains that range from 7.5$\%$ up to 58$\%$ for 64K (65536)-QAM when compared to Nyquist signaling. 
Additionally, for binary phase-shift keying (BPSK) and QPSK, the proposed ADMMSE FTN signaling detector significantly outperforms the successive symbol-by-symbol with go-back-\textit{K} sequence estimation (SSSgb\textit{K}SE) algorithms in \cite{bedeer2017very} at the expense of higher computational time. 

The remainder of this paper is organized as follows. Section \ref{recent} discusses the recent FTN signaling works. Section \ref{sys} presents the FTN signaling system model and formulates the detection problem. Section \ref{proposed} is dedicated to discuss the details of the proposed ADMMSE FTN signaling detection algorithm.  In Section \ref{Sim}, the simulation results are discussed and Section \ref{Conclude} concludes the paper.
\section{Recent FTN Progress}\label{recent}
 
The SE (in bits/sec/Hz) is the number of information bits carried per a given time and bandwidth, and it can be increased by increasing the transmission power (in addition to independently increasing the transmission bandwidth) to support high-order modulation. FTN signaling can be seen as an alternative or a complement to high-order modulation to further improve the SE. Surprisingly, a very similar concept is now the driving force behind holographic communication where researchers showed that for space-limited holographic signals it is possible to perfectly reconstruct holographic objects from their (significantly) below Nyquist rate samples \cite{4373328}. This shows the potential of improving the performance of communications systems by accepting interference in other applications and domains. 

After Mazo’s work, the concept of Mazo limit has been extended to different domains, i.e., other pulse shapes \cite{liveris2003exploiting}, high-order (up to 8-ary) pulse amplitude modulation \cite{380028,4524864}, and MIMO systems \cite{4801456} to name a few extensions. Most importantly, the concept of Mazo limit has been extended to the frequency domain where the spacing between the subcarriers of orthogonal frequency division multiplexing (OFDM) transmission is packed beyond the orthogonal condition and up to a certain limit. Such frequency packing results in intentional inter-carrier interference (ICI) while maintaining the same minimum Euclidean distance, and hence, the error rate of OFDM transmission \cite{rusek2005two} as long as there are proper detection techniques at the receiver to remove the ICI. Such non-orthogonal frequency domain transmission is usually called multi-stream FTN signaling \cite{4939227} or time-frequency packing \cite{5288497}, and it is considered a promising candidate waveform for next-generation communications systems. The importance of extending the concept of Mazo limit to the frequency domain is due to the fact that the independent SE gains can be obtained from time-domain pulse acceleration and frequency-domain subcarrier packing. 

The optimal detection algorithms of the FTN signaling that minimize the error rate in the presence of ISI and/or ICI are in general complex \cite{anderson2013faster}. Following \cite{anderson2013faster},  we define three orders of detector complexity: simple, trellis, and iterative, depending on the severity of the interference and how much processing power is available at the receiver for a given application. Simple detection can be in the form of simple equalization techniques to remove the ISI/ICI such as the works in \cite{bedeer2017very, 9120701, 6574905, ibrahim2021novel, 7510967}, where acceptable error rate performance is reported for light ISI scenarios, i.e., at values of time acceleration parameters around 0.9 or 0.8. 
Semi-definite relaxation techniques in \cite{bedeer2017low} and \cite{bedeer2019low} show that moderate gains of the SE can be obtained with polynomial time complexity for the detection of  high-order QAM and phase shift keying (PSK) modulations, respectively.
As the ISI length grows, simple detectors fail to give satisfactory error rate performance and more involved detectors are required. The ISI generated from FTN signaling has a trellis structure \cite{liveris2003exploiting} and techniques such as standard Viterbi algorithm (VA) or Bahl-Cocke-Jelinek-Raviv (BCJR) algorithm can be used to detect the most likely transmit sequence or to find the likelihood of individual bits, respectively, for moderate levels of ISI/ICI \cite{liveris2003exploiting, 5205622}. Iterative detection or turbo equalization is useful when the trellis size becomes large in severe ISI scenarios \cite{anderson2013faster}. In such cases, standard VA or BCJR become impractical to implement and either reduced trellis or reduced search variants of the VA or BCJR are used \cite{5205622, 4595029}, where the main idea is to either search a part of the whole trellis (reduced search); or to perform full search only over a smaller trellis size (reduced trellis). Such iterative techniques showed good performance for coded FTN signaling systems \cite{6241379}. 

{\color{black}Low-complexity detection of FTN signaling attracted the attention of the research community \cite{ibrahim2021novel, wen2020time, wen2019optimization, jana2018dual, jana2017pre, li2020code, li2017reduced, rusek2011optimal}. For instance, in \cite{ibrahim2021novel}, the authors  proposed a reduced complexity detector for FTN signaling based on the primal-dual predictor-corrector interior point algorithm (CQRAQSE) that can efficiently detect BPSK and QPSK FTN signaling. 
To further improve the SE, the authors in \cite{wen2020time} proposed a time-frequency compressed FTN signaling scheme that can improve the SE in two dimensions. In particular, the SE is improved in the time-domain by accelerating the transmit symbols beyond the Nyquist limit; while the SE improvement in the frequency domain is achieved by precoding. The authors in \cite{wen2019optimization} designed a precoding technique that resulted in introducing an artificial interference to help in the detection of FTN signaling in severe ISI scenarios. To support high-order modulation FTN signaling at reduced computational complexity, Tomlinson-Harashima precoding (THP) was used in \cite{jana2017pre} along with two novel soft demapping algorithms to compensate for the modulo-loss associated with the THP. In \cite{li2020code}, a new convolutional code is designed to absorb the channel memory caused by FTN signaling. Then, joint detection and decoding is designed based on the integrated trellis of the convolutional coded FTN signaling. In \cite{li2017reduced}, the M-algorithm BCJR is redesigned based on the Ungerboeck observation model to avoid having a whitening matched filter. }

Various other specific aspects of the FTN signaling technology have been recently under active research.  In \cite{8798843} a zero-forcing one tap equalizer and trellis demodulator of non-orthogonal multicarrier FTN signaling over LTE fading channels is studied. Joint FTN signaling detection, channel estimation, and user-activity tracking is investigated for an FTN-NOMA system relying on random access to support massive connectivity and high throughput in machine-type communications in dynamically fluctuating environments \cite{9006927}.  In \cite{9185013}, FTN signaling over frequency-selective fading channels is considered for which a joint channel estimation and precoding algorithm is proposed to perform data detection for FTN signaling. In comparison to the existing frequency-domain channel estimation and equalization algorithms, the algorithm in \cite{9185013} results in complexity reduction for the signal processing at receivers, thanks to the linear precoding processing on the transmitter's side. In \cite{8866862}, to avoid the need for a dedicated channel for control frames, a deep neural network architecture was proposed to estimate the adaptive time acceleration parameter. The estimation accuracy reached 99\% for acceleration parameters in the range of 0.6 to 1. 

Another important design problem in FTN signaling is synchronization. Due to the inherent ISI in the transmitted FTN signal, traditional synchronization methods become unsuitable and inaccurate making the carrier synchronization of FTN systems more involved. To solve this problem, an iterative carrier frequency offset  estimation scheme for an FTN signaling system is proposed in \cite{8301798} based on discrete Fourier transform and a golden-section search algorithm.

\begin{figure}[t!b!]
	\begin{center}
	\centering
\includegraphics[width=1.0\columnwidth]{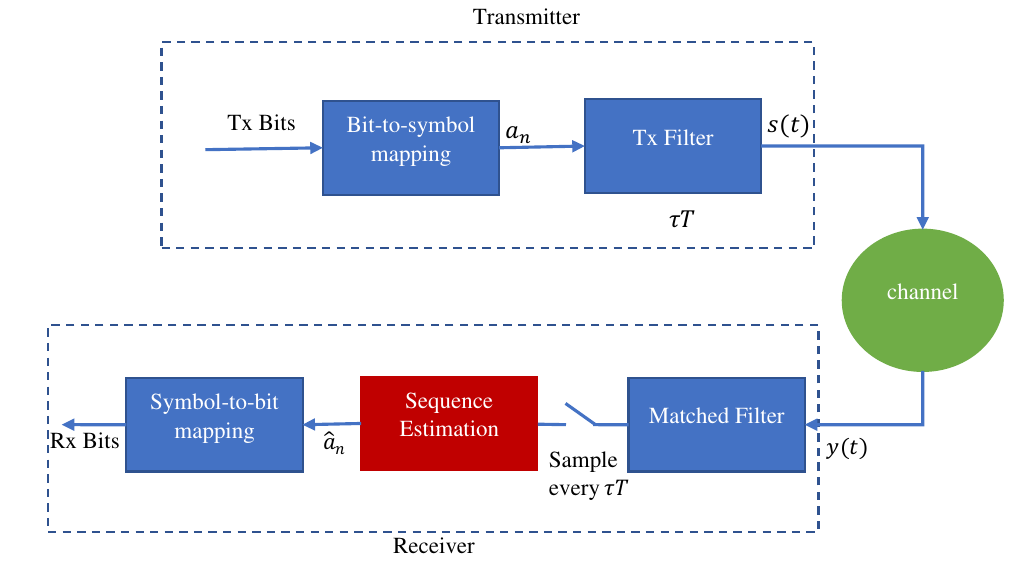}%
\caption{Block diagram of FTN signaling system.}
\label{system_model}
\end{center}
\end{figure}
\section {System Model} \label{sys}

Fig. \ref{system_model} shows the block diagram of a possible transmitter and receiver structure of a single carrier FTN signaling system. At the transmitter, the data bits to be transmitted are gray mapped to symbols at the bits-to-symbols mapping block. These data symbols are then transmitted through the transmit pulse shaping filter at a rate faster than Nyquist's, i.e., at $R_{s}=\frac{1}{\tau T}$, where $R_{s}$ is the signaling rate, $0<\tau \leq 1$ is the acceleration parameter, and $T$ is the orthogonal symbol time. The transmitted signal $s\left(t\right)$ of the FTN signaling system shown in Fig. \ref{system_model} can be written in the form
\begin{equation}
s\left(t\right)=\sum_{n=1}^{N}a_{n}p\left(t-n\tau T\right),~~~0<\tau < 1,
\end{equation}
where $N$ is the total number of transmit data symbols, $a_n$, is the independent and identically distributed data symbols drawn from the \textit{M}-QAM modulation constellation with energy $E_{s}$,  and $p\left(t\right)$ is a unit-energy pulse, i.e. $\int_{-\infty}^{\infty}\left|p\left(t\right)\right|^{2}dt=1$.

At the receiver, the FTN signal is given as
\begin{equation}
y\left(t\right)=s\left(t\right)+n\left(t\right),
\end{equation}
where $n(t)$ is a zero mean complex valued Gaussian random variable with variance $\sigma^2$. { In this paper, we consider communication systems that employ ultra high-order modulation, such as DOCSIS and microwave links, and hence, AWGN channel is considered as a suitable channel model rather than the frequency-selective fading channels encountered in other wireless systems.} At the first stage of the receiver, we use a filter matched to $p(t)$, and the received signal after the matched filter can be written as
\begin{equation}
\bar{y}\left(t\right)=\sum_{n=1}^{N}a_{n}g\left(t-n\tau T\right)+w\left(t\right),
\end{equation}
where $g\left(t\right)=p\left(t\right)\ast p\left(T-t\right)$, and $w\left(t\right)=n\left(t\right)\ast p\left(T-t\right)$.
The sampled received  FTN signal is expressed as
\begin{align*}\label{sampled}
\hat{y}_{k}&=\hat{y}\left(k\tau T\right)\\
&=\sum_{n=1}^{N}a_{n}g\left(k\tau T-n\tau T\right)+w\left(k\tau T\right)\\
&= \underset{\text{desired symbol}}{\underbrace{a_{n}g\left(0\right)}}+\underset{\text{ISI}}{\underbrace{ \sum_{n=1,n \neq k}^{N}a_{n}g\left(\left(k-n\right)\tau T\right)}}+w\left(k\tau T\right).\numberthis 
\end{align*}
Equation (\ref{sampled}) shows that  for a given \textit{k}th received symbol, there are components from 
the \textit{k}th transmitted symbol, as well as ISI from adjacent symbols.  This can be re-written in vector form as
\begin{equation}\label{FTN_sig}
\mathbf{y}^{c}=\mathbf{G}\mathbf{a}+\mathbf{w}^{c},
\end{equation}
where $\mathbf{y}^{c}$ is the complex column vector of received samples of size $N$,  $\mathbf{a}$ is the transmitted data symbols vector of length $N$, and \textbf{G} is the $N \times N$ intersymbol interference (ISI) matrix, which is Toeplitz and symmetric. Instead of working with complex numbers, we re-write (\ref{FTN_sig}) as
\begin{align*}\label{rec_signal1}
\tilde{\mathbf{y}}_{2 N\times1}&=\tilde{\mathbf{G}}_{2 N \times 2N}\tilde{\mathbf{a}}+\tilde{\mathbf{w}}_{2 N\times1},\\
\begin{bmatrix}
\mathcal{R}\left(\mathbf{y}^{c}\right)\\ 
\mathcal{I}\left(\mathbf{y}^{c}\right)
\end{bmatrix}&=\begin{bmatrix}
\mathbf{G}& \mathbf{0}_{N \times N}\\ 
\mathbf{0}_{N \times N} &  \mathbf{G}
\end{bmatrix}\begin{bmatrix}
\mathcal{R}\left(\mathbf{a}\right) 
\\ 
\mathcal{I}\left(\mathbf{a}\right) 
\end{bmatrix}+\begin{bmatrix}
\mathcal{R}\left(\mathbf{w}^{c}\right) 
\\ 
\mathcal{I}\left(\mathbf{w}^{c}\right) \numberthis
\end{bmatrix}.
\end{align*}
As concluded in \cite{bedeer2017reduced}, $\mathbf{w}^{c}\sim \mathcal{N}\left(0,\sigma^{2}\mathbf{G}\right)$ , which indicates that the noise samples at the sampler's output are Gaussian distributed with non-diagonal co-variance matrix. This means that the elements of the noise vector $\mathbf{w}^{c}$ are correlated. 

Following \cite{johnson2013statistical, 4595029}, one can show that the ISI matrix $\mathbf{G}$ and the matrix $\tilde{\mathbf{G}}$ will be positive-definite, and hence, invertible for general pulse shapes. In this case, we could express  {(\ref{rec_signal1})} as
\begin{align*}\label{z}
\tilde{\mathbf{G}}^{-1}\tilde{\mathbf{y}}&=\tilde{\mathbf{a}}+\tilde{\mathbf{G}}^{-1}\tilde{\mathbf{w}},\\
\mathbf{z}&=\tilde{\mathbf{a}}+\mathbf{n},\numberthis
\end{align*}
where $\mathbf{z}=\tilde{\mathbf{G}}^{-1}\tilde{\mathbf{y}}$ and $\mathbf{n}=\tilde{\mathbf{G}}^{-1}\tilde{\mathbf{w}}$. The FTN signaling detection problem maximizes the probability that the transmit symbols $\tilde{\mathbf{a}}$ are sent given the received samples $\mathbf{z}$. By invoking Bayes theorem, an equivalent expression is hence given by
\begin{equation}\label{ML}
\tilde{\mathbf{a}}=\underset{\tilde{\mathbf{a}}\in \mathcal{D}}{\max}p\left(\mathbf{z}|\tilde{\mathbf{a}}\right),
\end{equation}
where $\mathcal{D}$ is the set of discrete levels for both the in-phase and quadrature components of the symbols, whose values depend on the modulation type and order. The likelihood probability $p\left(\mathbf{z}|\tilde{\mathbf{a}}\right)$ needs to be maximized in order for the detection to be optimal. The problem of finding an estimate vector $\hat{\mathbf{a}}$ that maximizes this probability is the maximum likelihood sequence estimation (MLSE) problem. The received samples $\mathbf{z}$ can be seen as Gaussian random variables with a mean $\tilde{\mathbf{a}}$ and covariance matrix $\frac{1}{2}\sigma^{2}\tilde{\mathbf{G}}^{-1}$ \cite{bedeer2017low}. The likelihood probability in (\ref{ML}) is written as \cite{johnson2013statistical}
\begin{equation}
p\left(\mathbf{z}|\tilde{\mathbf{a}}\right)=\left(\frac{1}{2\pi \sigma^{2}}\right)^{\frac{N}{2}}\frac{1}{\sqrt{\text{det}\left[\tilde{\mathbf{G}}^{-1}\right]}}\text{e}^{-\frac{1}{\sigma^{2}}\left(\mathbf{z}-\tilde{\mathbf{a}}\right)^{\top }\tilde{\mathbf{G}}\left(\mathbf{z}-\tilde{\mathbf{a}}\right)}.
\end{equation}
We can therefore formulate the MLSE problem for detecting the FTN signal in (\ref{FTN_sig}) as \cite{bedeer2017low}
\begin{equation}\label{prob1}
\hat{\mathbf{a}}=\arg ~ \underset{\tilde{\mathbf{a}}\in \mathcal{D}}{\min}\left(\mathbf{z}-\tilde{\mathbf{a}}\right)^{\top }\tilde{\mathbf{G}}\left(\mathbf{z}-\tilde{\mathbf{a}}\right).
\end{equation}

For high FTN signaling rates, the FTN signaling has spectral zeros \cite{4595029} and the ISI matrix could run into conditioning issues. In this case, rather than having to invert the matrix as in (\ref{z}), the received symbol vector $\tilde{\mathbf{y}}$ is passed through an approximate noise whitening filter \cite{4595029} that can be derived from spectral factorization. We can then re-write  (\ref{FTN_sig}) as 
\begin{equation}\label{conv}
\mathbf{y}^{c}= \mathbf{a}\ast  \mathbf{g}+\mathbf{w}^{c},
\end{equation}
where $\ast$ is the convolution operator. 
Hence, after passing (\ref{conv}) through the approximate whitening filter, we have
\begin{equation}\label{eq:eb}
\mathbf{y}^{uncorr}=\mathbf{a} \ast \mathbf{v}+\mathbf{w}^{uncorr},
\end{equation}
where $\mathbf{w}^{uncorr}$ is white Gaussian noise with zero mean and variance $\sigma^{2}$ and $\mathbf{v}$ represents the causal ISI such that $\mathbf{v}\left[n\right] \ast \mathbf{v}\left[-n\right]=\mathbf{g}$. We can then re-write (\ref{eq:eb}) in a vector form as 
\begin{equation}
\mathbf{y}^{uncorr}=\mathbf{V}\mathbf{a}+\mathbf{w}^{uncorr},
\end{equation}
where $\mathbf{V}$ is a Gram Toeplitz matrix, and hence, is positive semi-definite. Using the equivalent real-valued model
\begin{align*}\label{rec_signal1_white}
\tilde{\mathbf{y}}_{2 N\times1}^{uncorr}&=\tilde{\mathbf{V}}_{2 N \times 2N}\tilde{\mathbf{a}}+\tilde{\mathbf{w}}_{2 N\times1}^{uncorr},\\
\begin{bmatrix}
\mathcal{R}\left(\mathbf{y}^{uncorr}\right) 
\\ 
\mathcal{I}\left(\mathbf{y}^{uncorr}\right) 
\end{bmatrix}&=\begin{bmatrix}
\mathbf{V}& \mathbf{0}_{N \times N}\\ 
\mathbf{0}_{N \times N} &  \mathbf{V}
\end{bmatrix}\begin{bmatrix}
\mathcal{R}\left(\mathbf{a}\right) 
\\ 
\mathcal{I}\left(\mathbf{a}\right) 
\end{bmatrix}+\begin{bmatrix}
\mathcal{R}\left(\mathbf{w}^{uncorr}\right) 
\\ 
\mathcal{I}\left(\mathbf{w}^{uncorr}\right) \numberthis
\end{bmatrix},
\end{align*}
and similar to the earlier discussion, we formulate the high-order QAM FTN signaling detection problem as
\begin{equation}\label{prob2}
\hat{\mathbf{a}}=\arg ~ \underset{\tilde{\mathbf{a}}\in \mathcal{D}}{\min}\left \| \tilde{\mathbf{y}}_{2 N\times1}^{uncorr}- \tilde{\mathbf{V}}_{2 N \times 2N}\tilde{\mathbf{a}}\right \|_{2}^{2}.
\end{equation}

Both FTN signaling detection problems in (\ref{prob1}) and (\ref{prob2}) are NP hard problems. Specifically speaking, the computational time required to optimally detect the received FTN signaling sequence in an AWGN channel is expected to grow exponentially with respect to the received sequence block length. In the next section, we propose the use of an optimization algorithm that exploits the positive semi-definite structures of $\mathbf{G}$ and $\mathbf{V}^{\top} \mathbf{V}$ to strike a balance between the computational complexity require for FTN signaling detection and the goodness of the sub-optimal solution (i.e. the performance). The algorithm is a variant of the alternating directions multiplier method (ADMM) modified to return good quality sub-optimal solutions for non-convex quadratic programs at low computational cost. The concept of operation for the ADMM is thoroughly detailed in the next section before summarizing the variant that we use for the ultra high-order QAM FTN signaling detection.

\section{The Proposed ADMM based FTN Signaling Detection}\label{proposed}

In any communication system, including FTN signaling, detection is done in real time as the symbols are received in blocks of length $N$. The computational power needed for the detection affects the complexity and latency, and hence, the cost of the receiver. Therefore, methods to find the global solution for the optimization problems in (\ref{prob1}) or (\ref{prob2}), if available, are not favorable because their large run-time cannot be tolerated. Hence, the optimization scheme discussed in this section is a simple, but powerful and computationally efficient routine to find high-quality approximate solutions to (\ref{prob1}) and (\ref{prob2}) quickly. In the jargon of mathematical programming, an algorithm that is not guaranteed to obtain the global optimum in a finite number of steps is known as a \textit{heuristic}. In this section, we discuss the mechanism of the proposed ADMM based detection scheme and its details which are employed for the FTN signaling detection problem. It is a variant of the ADMM, which is an optimization algorithm founded on two main concepts, which are the \textit{augmented Lagrangian methods} \cite{birgin2014practical} and the \textit{method of multipliers} \cite{bertsekas2014constrained}, both from the field of convex optimization which are detailed out in Section \ref{Aug}. In Section \ref{prem}, we discuss how the original ADMM proposed for convex optimization problems works and how it guarantees obtaining the global minimizer for these problems.  In Section \ref{ADMMSE}, we unroll the modified ADMM  \cite{takapoui2020simple,boyd2011distributed}  as a heuristic method for solving the quadratic non-convex FTN signaling detection problems in equations (\ref{prob1}) and (\ref{prob2}) to obtain good quality sub-optimal solutions, and hence, achieve high performance for the FTN signaling detection. 

\subsection{Augmented Lagrangian Methods and the Method of Multipliers}\label{Aug}

Augmented Lagrangian methods bring robustness to the dual descent method \cite{luo1993convergence}. Consider the equality-constrained convex optimization problem
\begin{align*} \label{1}
~~~~~~~~~~&\underset{\mathbf{x}}{\min }~f\left(\mathbf{x}\right) \\
&\text{subject to}~~\mathbf{A}\mathbf{x}=\mathbf{b}\numberthis,
\end{align*}
where $\mathbf{x}\in\mathbb{R}^{n}$, $\mathbf{b} \in \mathbb{R}^{m}$, $\mathbf{A} \in \mathbf{R}^{m \times n}$ and $f:\mathbb{R}^{n}\rightarrow \mathbb{R}$ is convex. The augmented Lagrangian for ($\ref{1}$) is \cite{boyd2011distributed}
\begin{equation} \label{Lag_aug}
\mathcal{L}_{\rho}\left(\mathbf{x},\bm{\lambda}\right)=f\left(\mathbf{x}\right)+\bm{\lambda}^{\top }\left(\mathbf{A}\mathbf{x}-\mathbf{b}+\frac{\rho}{2}\left \| \mathbf{Ax}-\mathbf{b} \right \|_{2}^{2}\right),
\end{equation}
where $\bm{\lambda}$ is the vector of Lagrangian multipliers, and $ \rho> 0$ is called the penalty parameter. The augmented Lagrangian can be viewed as the (unaugmented) Lagrangian associated with the problem
\begin{align*} \label{2}
~~~~~~~~~~&\underset{\mathbf{x}}{\min }~\left(f\left(\mathbf{x}\right)+\frac{\rho}{2}\left \| \mathbf{Ax}-\mathbf{b} \right \|_{2}^{2}\right) \\
&\text{subject to}~~\mathbf{A}\mathbf{x}=\mathbf{b}, \numberthis
\end{align*}
which is equivalent to the problem in (\ref{1}) since for any feasible $\mathbf{x}$ the term added to the objective is zero. The associated
dual function is $g_{\rho}\left(\bm{\lambda}\right)=\underset{\mathbf{x}}{\min }\left(\mathcal{L}_{\rho}\left(\mathbf{x},\bm{\lambda}\right)\right)$.

The gradient of the augmented dual function is computed by minimizing over $\mathbf{x}$, and then evaluating the resulting equality constraint residual. Applying dual ascent yields
\begin{subequations}
\begin{align}
\mathbf{x}^{k+1} & := arg~\underset{\mathbf{x}}{\min } \mathcal{L}_{\rho}\left(\mathbf{x},\bm{\lambda}^{k}\right) \\
\bm{\lambda}^{k+1}& :=\bm{\lambda}^{k}+\rho\left(\mathbf{A}\mathbf{x}^{k+1}-\mathbf{b}\right),
\end{align}
\end{subequations}
 which is the method of multipliers. The method of multipliers requires much less stringent conditions to converge compared to the dual ascent. By using $\rho$ as the step size in the dual update, the iterate $\left(\mathbf{x}^{k+1},\bm{\lambda}^{k+1}\right)$ is dual feasible. As the method of multipliers progresses, the primal residual $\mathbf{A}\mathbf{x}^{k+1}-\mathbf{b}$ converges to zero and hence, optimality.

\subsection{The Concept of ADMM}\label{prem}
ADMM is a procedure that coordinates decomposition where the solutions to small local sub-problems are coordinated to find a solution to a large global problem \cite{boyd2011distributed}. It combines the benefits of dual decomposition with augmented Lagrangian methods for constrained optimization. The algorithm solves problems in the form
\begin{align*} \label{2}
~~~~~~~~~~&\underset{\mathbf{x}_{1},\mathbf{x}_{2}}{\min }~f_{1}\left(\mathbf{x}_{1}\right)+f_{2}\left(\mathbf{x}_{2}\right) \\
&\text{subject to}:~~\mathbf{A}_{1}\mathbf{x}_{1}+\mathbf{A}_{2}\mathbf{x}_{2}=\mathbf{b}, \numberthis
\end{align*}
where $\mathbf{x}_1 \in \mathbb{R}^{n_1}$, $\mathbf{x}_2 \in \mathbb{R}^{n_2}$, $\mathbf{A}_{1} \in \mathbb{R}^{p \times n_{1}}$, $\mathbf{A}_{2} \in \mathbb{R}^{p\times n_{2}}$, and $\mathbf{b} \in \mathbb{R}^{p}$. The augmented Lagrangian is formed as 
\begin{align}
\mathcal{L}_{\rho}\left(\mathbf{x}_{1},\mathbf{x}_{2},\bm{\lambda}\right)=&f_{1}\left(\mathbf{x}_{1}\right)+f_{2}\left(\mathbf{x}_{2}\right)+\bm{\lambda}^{\top }\left(\mathbf{A}_{1}\mathbf{x}_{1}+\mathbf{A}_{2}\mathbf{x}_{2}-\mathbf{b}\right) \nonumber \\
&+\left(\frac{\rho}{2}\right)\left \| \mathbf{A}_{1}\mathbf{x}_{1}+\mathbf{A}_{2}\mathbf{x}_{2}-\mathbf{b} \right \|^{2}_{2}.
\end{align}
The ADMM consists of the iterations
\begin{subequations}
\begin{align}
\mathbf{x}_{1}^{k+1} & := arg~\underset{\mathbf{x}_1}{\min } \mathcal{L}_{\rho}\left(\mathbf{x}_{1},\mathbf{x}_{2}^{k},\bm{\lambda}^{k}\right), \\
\mathbf{x}_{2}^{k+1} & := arg~\underset{\mathbf{x}_2}{\min } \mathcal{L}_{\rho}\left(\mathbf{x}_{1}^{k+1},\mathbf{x}_{2},\bm{\lambda}^{k}\right), \\
\bm{\lambda}^{k+1}& :=\bm{\lambda}^{k}+\rho\left(\mathbf{A}_{1}\mathbf{x}_{1}^{k+1}+\mathbf{A}_{2}\mathbf{x}_{2}^{k+1}-\mathbf{b}\right),
\end{align}
\end{subequations}
where $\rho>0$. We update the dual variable with a step size equal to $\rho$, similar to the update in the method of multipliers.
While in the method of multipliers the augmented Lagrangian is minimized jointly with respect to
the two primal vectors, in ADMM on the other hand, $\mathbf{x}_{1}$ and $\mathbf{x}_{2}$ are updated in an alternating or sequential fashion, which accounts for the term alternating directions. 

We combine the linear and quadratic terms in the augmented Lagrangian and scale the dual variable to obtain an equivalent representation of the ADMM. By defining the residual $\bm{\gamma}=\mathbf{A}_{1}\mathbf{x}_{1}+\mathbf{A}_{2}\mathbf{x}_{2}-\mathbf{b}$, we have
\begin{align}
\bm{\lambda}^{\top }\bm{\gamma}+\frac{\rho}{2}\left \|\bm{\gamma}  \right \|_{2}^{2}~=~&\frac{\rho}{2}\left \| \bm{\gamma} + \frac{1}{\rho}\bm{\lambda} \right \|_{2}^{2}-\frac{1}{2\rho}\left \| \lambda \right \|_{2}^{2}, \nonumber \\
~=~&\frac{\rho}{2}\left \| \bm{\gamma} +\bm{\mu}\right \|_{2}^{2}-\left(\frac{\rho}{2}\right)\left \|  \bm{\mu}\right \|_{2}^{2}, 
\end{align}
where $\bm{\mu}=\frac{1}{\rho}\bm{\lambda}$ is the scaled dual variable vector. Hence, the ADMM can then be expressed as

\begin{subequations}
\begin{align}
\mathbf{x}_{1}^{k+1} & := arg~\underset{\mathbf{x}_1}{\min }\left(f\left(\mathbf{x}_{1}\right)+\frac{\rho}{2}\left \| \mathbf{A}_{1}\mathbf{x}_{1}+  \mathbf{A}_{2}\mathbf{x}_{2}^{k} -\mathbf{b}+\bm{\mu}^{k}\right \|_{2}^{2}\right)\label{eq_1}, \\
\mathbf{x}_{2}^{k+1} & :=  arg~\underset{\mathbf{x}_2}{\min }\left(g\left(\mathbf{x}_{2}\right)+\frac{\rho}{2}\left \| \mathbf{A}_{1}\mathbf{x}_{1}^{k+1}+  \mathbf{A}_{2}\mathbf{x}_{2} -\mathbf{b}+\bm{\mu}^{k}\right \|_{2}^{2}\right) \label{eq_2}, \\
\bm{\mu}^{k+1}& :=\bm{\mu}^{k}+\mathbf{A}_{1}\mathbf{x}_{1}^{k+1}+\mathbf{A}_{2}\mathbf{x}_{2}^{k+1}-\mathbf{b} \label{eq_3}.
\end{align}
\end{subequations}

ADMM can also be exploited for nonconvex problems like (\ref{prob1}) and (\ref{prob2}). In this case, it does not have to converge to an optimal point. It can converge to different (and in particular, nonoptimal) points, depending on the initial values $\mathbf{x}^{0}$ and $\bm{\mu}^{0}$ and the parameter $\rho$. In the next sub-section, we discuss how ADMM is used for the non-convex FTN signaling detection problem in (\ref{prob1}).

\subsection{ADMM Sequence Estimation (ADMMSE) FTN Signaling Detector}\label{ADMMSE}

 The problem in (\ref{prob1}) can be reformulated as
\begin{equation} \label{op}
\hat{\mathbf{a}}=\arg ~\left(\underset{\tilde{\mathbf{a}}\in \mathcal{D}}{\min }~\tilde{\mathbf{a}}^{\top }\tilde{\mathbf{G}}\tilde{\mathbf{a}}+\mathbf{q}^{\top }\tilde{\mathbf{a}}+r\right),
\end{equation}
where $\mathbf{q}^{\top }=-2\mathbf{z}^{\top }\tilde{\mathbf{G}}$ and $r=\mathbf{z}^{\top }\tilde{\mathbf{G}}\mathbf{z}$. In a similar manner, the optimization problem in (\ref{prob2}) can be reformulated as
\begin{equation}\label{opt2b}
\hat{\mathbf{a}}=\arg ~\left(\underset{\tilde{\mathbf{a}}\in \mathcal{D}}{\min }~\tilde{\mathbf{a}}^{\top }\mathbf{H}\tilde{\mathbf{a}}+\mathbf{q}^{\top }_{uncorr}\tilde{\mathbf{a}}+r_{uncorr}\right),
\end{equation}
where $\mathbf{H}=\tilde{\mathbf{V}}^{\top}\tilde{\mathbf{V}}$ is a positive semi-definite matrix, $\mathbf{q}^{\top}_{uncorr}=-2\left(\tilde{\mathbf{y}}_{2 N\times1}^{uncorr}\right)^{\top}\tilde{\mathbf{V}}$, and $r_{uncorr}=\left(\tilde{\mathbf{y}}_{2 N\times1}^{uncorr}\right)^{\top}\tilde{\mathbf{y}}_{2 N\times1}^{uncorr}$.
For this type of optimization problems, there are many proposed sub-optimal techniques, one of the recent ones is is that proposed by R. Takapoui \textit{et al}. in \cite{takapoui2020simple} in the past few years. It is meant to solve non-convex quadratic optimization problems where the objective function is convex, if the discrete requirements on the decision variables are dropped. It is based on a variant of the ADMM algorithm  described earlier. Because (\ref{op}) and (\ref{opt2b}) fall under the non-convex optimization problems class, the proposed ADMM variant for the FTN signaling detection non-convex problem may not achieve the global optimal. While ADMM was introduced initially for convex optimization problems \cite{boyd2011distributed}, it turns out to be a promising candidate technique to find sub-optimal solutions for some NP-hard nonconvex problems, such as the FTN signaling detection.

We rewrite problem (\ref{op})\footnote{The following steps apply to the problem in (\ref{opt2b}) as well.} after adding the auxiliary vector $\mathbf{x} \in\mathbb{R}^{2N \times 1}$  as 

\begin{align*}
&~~~~~~~~\min ~\tilde{\mathbf{a}}^{\top }\tilde{ \mathbf{G}}\tilde{\mathbf{a}}+\mathbf{q}^{\top }\tilde{\mathbf{a}}+r+I_{\mathcal{D}}\left(\mathbf{x}\right) \\
&\text{subject to}~~\tilde{\mathbf{a}}-\mathbf{x}=\mathbf{0}_{2N \times 1}, \numberthis
\end{align*}
where $I_{\mathcal{D}}\left(\mathbf{x}\right)$ denotes the penalty function of $\mathcal{D}$, such that $I_{\mathcal{D}}\left(\mathbf{x}\right)=0$ for $\tilde{\mathbf{a}} \in \mathcal{D}$ and $I_{\mathcal{D}}\left(\mathbf{x}\right)= \infty$ for $\tilde{\mathbf{a}} \notin \mathcal{D}$. Each
iteration in the algorithm consists of the following three steps
\begin{subequations}
\begin{align}
\tilde{\mathbf{a}}^{k+1}:=&\arg \underset{\tilde{\mathbf{a}}}{\min}\left(f\left(\tilde{\mathbf{a}}\right)=\tilde{\mathbf{a}}^{\top } \tilde{\mathbf{G}}\tilde{\mathbf{a}}+\mathbf{q}^{\top }\tilde{\mathbf{a}}+r+\frac{\rho}{2}\left \| \tilde{\mathbf{a}}- \mathbf{x}^{k}+\bm{\mathbb{\mu}}^{k}\right \|_{2}^{2}\right), \label{ADMM_eq1}\\
\mathbf{x}^{k+1}:=&\Xi \left(\tilde{\mathbf{a}}^{k+1}+\bm{\mathbb{\mu}}^{k}\right),  \label{ADMM_eq2}\\
\bm{\mathbb{\mu}}^{k+1}:=&\bm{\mathbb{\mu}}^{k}+\tilde{\mathbf{a}}^{k+1}-\mathbf{x}^{k+1}.  \label{ADMM_eq3}
\end{align}
\end{subequations}
Here, $\Xi \left(\cdot\right)$ represents a projection on the set $\mathcal{D}$, the vector $\bm{\mathbb{\mu}}\in \mathbb{R}^{2N}$ is the vector of multipliers and $\rho \in \mathbb{R}$ is a scalar. The projection $\Xi \left(\mathbf{x}\right) \in \arg \underset{\tilde{\mathbf{a}}\in\mathcal{D}}{\min}\left \| \tilde{\mathbf{a}}-\mathbf{x} \right \|_{2},~\forall~\mathbf{x} \in \mathbb{R}^{2N}$. Since $\mathcal{D}$ is the cartesian product of subsets of the real line, i.e., $\mathcal{D}=\mathcal{D}_{1} \times \cdots \times \mathcal{D}_{2N}$ (in our detection problem discrete levels), then we can consider $\Xi \left(\mathcal{\mathbf{x}}\right)=\bigg( \Xi_{1} \left(x_{1}\right) , \cdots ,  \Xi_{2N} \left(x_{2 N} \right)\bigg)$, where $\Xi_{i} $ is a projection function on to $\mathcal{D}_{i}$. Since $\mathcal{D}_{i}$ is a set of discrete values (integers for QAM), $\Xi_{i}$ rounds its argument to the nearest feasible discrete level. For any finite set $\mathcal{D}_{i}$ with $m$ elements, $\Xi_{i}\left({x_{i}}\right)$ is the closest point to $x_{i}$ that belongs to $\mathcal{D}_{i}$ which can be found by $\left \lceil \log_{2}m \right \rceil$ comparisons \cite{takapoui2020simple}. For a given modulation scheme, $m$ is the larger number of discrete levels of either the in-phase or the quadrature components. For a QAM modulation with a square constellation and order $M$, $m= \sqrt{M}$. {  It is worthy to emphasize that for the proposed ADMMSE, the modulation order impacts only the  step of quantization which turns out to be the least expensive step in terms of the computational effort (requires $\left \lceil \log_{2} \sqrt{M} \right \rceil$ comparison operations). Such low sensitivity of the computational complexity to the modulation order makes the proposed ADMMSE a promising candidate to detect ultra high-order QAM FTN signaling.}

The most computationally expensive step is the first step in (\ref{ADMM_eq1}), which involves minimizing an unconstrained strongly convex multivariate quadratic function. For such a problem, it is well known from mathematical optimization fundamentals, that a local minimizer is also a unique global minimizer. This unique global minimizer necessarily has $\nabla f\left(\tilde{\mathbf{a}}\right)=\mathbf{0}$, where $\nabla$ is the gradient operator. Given that this is a quadratic multivariate function, the gradient would result in a set of linear equations that has a unique solution. It is easy to show that the linear system that results is
\begin{equation}\label{lin_sys}
\left [ \mathbf{\tilde{\mathbf{G}}}+ \rho I_{2N \times 2N} \right ]\tilde{\mathbf{a}}^{k+1}=\mathbf{c},
\end{equation}
where $\mathbf{c}=-\mathbf{q}+\rho\left(\mathbf{x}^{k}-\bm{\mu}^{k}\right)$. 
The initialization of $\tilde{\mathbf{a}}^{0}$ is simply done by choosing a random point in the convex hull relaxation, $\mathbf{Co}~\mathcal{D}$, of the discrete lattice. Based on our experience with the numerical FTN signaling detection simulations, it turns out that running the algorithm more than once with different random initializations increases the chance of finding better quality solutions, and hence leads to better performance. The initial value of the multiplier vector $\bm{\mu}^{0}$ is set to $\mathbf{0}$. Moreover, the discussions in \cite{takapoui2020simple} suggest that the precision and convergence rate of first-order methods can be significantly improved by preconditioning the problem. The preconditioning used in our detector simply divides $\tilde{\mathbf{G}}$ and $\mathbf{q}$ by the maximum singular value of $\tilde{\mathbf{G}}$. Finally it is worth mentioning that since the objective value need not decrease monotonically, it is critical to keep track of the best point found. The ADMMSE FTN detection algorithm is summarized in Algorithm \ref{ADMM}, {where $precondition\left(\right)$ and $factorize\left(\right)$ are the preconditioning and $LDL^{\top}$ factorization, respectively. $linsolve\left(\right)$ is the subroutine responsible for solving the linear system in (\ref{lin_sys}).}


    \begin{algorithm}
        \caption{{Proposed ADMMSE FTN Signaling Detection Algorithm}\label{ADMM}}
        \begin{algorithmic}[1]\label{ADMM}
            \STATE \textbf{Input:} Pulse shape $p\left(t\right)$, ISI matrix $\mathbf{G}$ or causal ISI matrix $\mathbf{V}$, received samples $\mathbf{y}$, $\mathcal{D}$, number of iterations $L$, and number of initialization $\kappa$.
            \STATE $\left(\tilde{\mathbf{G}},\mathbf{q}\right)$ $\gets precondition\left(\tilde{\mathbf{G}},\mathbf{q}\right)$
            \STATE$\left(\mathbf{L},\mathbf{D}\right)\gets factorize(\tilde{\mathbf{G}})$ 
            \STATE$\tilde{\mathbf{a}}_{best} \gets \varnothing$ , $f\left(\tilde{\mathbf{a}}_{best}\right)\gets\infty$
            \FOR{Random initialization $1,2,\cdots, \kappa$}
                \FOR{Iteration $1,2,\cdots,L$} 
                    \STATE $\tilde{\mathbf{a}}^{k+1} \gets$ linsolve$\left(\mathbf{L},\mathbf{D},\mathbf{c}, \rho\right)$
                    \STATE${\mathbf{x}^{k+1}\gets  \Xi \left(\tilde{\mathbf{a}}^{k+1}+\bm{\mathbb{\mu}}^{k}\right)}$\\
                    \STATE${\bm{\mathbb{\mu}}^{k+1}\gets  \bm{\mathbb{\mu}}^{k}+\tilde{\mathbf{a}}^{k+1}-\mathbf{x}^{k+1}}$\\
                    \IF{$f\left(\tilde{\mathbf{a}}\right)<f\left(\tilde{\mathbf{a}}_{best}\right)$} 
                        \STATE {$\tilde{\mathbf{a}}_{best}\gets\tilde{\mathbf{a}}$} 
                    \ENDIF
                \ENDFOR
            \ENDFOR
        \RETURN  {$\tilde{\mathbf{a}}_{best}$}
        \end{algorithmic}
    \end{algorithm}


\subsection{Complexity of ADMMSE}
The most computationally expensive step in the ADMMSE is that required to solve the linear system in (\ref{lin_sys}). This can be solved at a worst case complexity of $\mathcal{O}\left(4N^{2}\right)$ when the coefficient matrix is factorized using $LDL^{\top }$ factorization \cite{takapoui2020simple}. The factorization process requires $\mathcal{O}\left(8N^{3}\right)$ \cite{takapoui2020simple}, however it is performed only once for AWGN channels, before any detection takes place and its result is cached and used through the detection process for every group of $N$ symbols. Therefore, when this gets penalized over a very large number of received blocks of $N$ symbols, its effect becomes negligible. Finally, the step (\ref{ADMM_eq3}) is simply an update step that involves just $4N$ additions and subtractions.  If the maximum number of iterations is $L$ and the number of initializations is $\kappa$, the overall algorithm complexity is $\mathcal{O}\left(4L\kappa N^{2}b_{1}+2NL\kappa \left \lceil \log_{2}m \right \rceil b_{2} + 4NL \kappa b_{3}\right)$, where $b_{1}$, $b_{2}$ and $b_{3}$ are the computational times, in seconds,  for each of (\ref{eq_1}-\ref{eq_3}), respectively, when $N=1$.   {As one can see, the computational effort of the ADMMSE scales with the logarithm of the square root of the modulation order. This makes the ADMMSE a promising candidate for the detection of ultra high-order QAM FTN signaling.}

{{It is important to note that in ADMMSE, the modulation order impacts only the quantization step which is the least expensive step, per iteration, in terms of computational effort. On the other hand, GASDRSE formulates the modulation constrains as a polynomial whose order increases with increasing the constellation alphabet leading to a significant increase in the SDR program's matrix size. The complexity of GASDRSE as shown in \cite{bedeer2017low} is $\mathcal{O}\left(\left(4N+1\right)^{3.5}+\left(4N+1\right)^{2}L_{rnd}\right)$, where $L_{rnd}$ is the number Gaussian randomization iterations. For the SSSgb$K$SE, it was shown in \cite{bedeer2017very} that for a given $K$, the complexity scales linearly in the ISI length, i.e., $\mathcal{O}\left(L_{ISI}\right)$, where $L_{ISI}$ is the ISI length, which depends on both $\tau$ and $\beta$. The computations require exactly $K\left(L_{ISI}-2\right)+\frac{K\left(K-1\right)}{2}$ additions and $K\left(L_{ISI}-1\right)+\frac{K\left(K+1\right)}{2}$ multiplications. Frequency domain equalization (FDE) FTN signaling detectors in \cite{6574905} and \cite{7510967}, as well as, the CPS-GTMH-DFTP linear FTN signaling precoder in \cite{{8954896}} have a complexity that scales as $\mathcal{O}\left(N\log\left(N\right)\right)$. The M-BCJR algorithm is known to have a good performance
if the state reduction ratio $\frac{2^{M L_{ISI}}}{M_{BCJR}}$ is not very large. However, it has to find the $M_{BCJR}$ largest elements in $M_{BCJR}2^{M}$ at every trellis section, which is computationally expensive to do for every state \cite{1523702}.}}

\subsection{Generation of Soft-Outputs}
To generate soft-outputs for channel coding, the ADMMSE can be used to decrease the complexity required to find the log-likelihood ratio (LLR) as follows. 
Using the idea of the list sphere decoding in \cite{1194444}, the ADMMSE can store a set of vectors returned after $L$ iterations for each of the $\kappa$ initial points. The set of vectors returned are the $\kappa$ vectors with the lowest objective function for either (\ref{prob1}) or (\ref{prob2}) rather than storing just one vector that achieves the smallest objective value that the ADMMSE finds after $\kappa L$ iterations. This set of selected vectors can be used in calculating an approximate value for the LLR as in the list sphere decoding in \cite{1194444}. Instead of searching the whole search space, ADMMSE efficiently produces candidate vectors that contribute the most towards the calculation of the LLR values.

\section{Simulation Results}\label{Sim}
In this section, the performance (in terms of BER and SE gains), as well as the computational time of our proposed ADMMSE detection algorithm for QPSK, 16-QAM, 64-QAM, 256-QAM, 1024-QAM, 4096-QAM, 16K (16,384)-QAM, and   64K (65,536)-QAM are evaluated. An rRC transmit pulse shaping filter is employed and the receiver is equipped with a matched filter, i.e., an rRC filter of the same roll-off factor as the transmit filter. The roll-off factors considered are $\alpha=0.3$ and $\alpha=0.5$, and the data symbols sequence length is $N=150$. The ADMMSE's parameter $\rho$ is numerically optimized for each modulation order. The ADMMSE is compared against the GASDRSE algorithm, which is based on the semi-definite relaxation (SDR) and Gaussian randomization, proposed in \cite{bedeer2019low,bedeer2017low}. For QPSK FTN signaling, the number of Gaussian randomization iterations is $J=10,000$. This number is chosen to be equal  to the aggregate number of iterations of  ADMMSE. For 16-QAM, we set $J=1000$ as in \cite{bedeer2017low}. For modulation orders greater than 16, we are not able to compare with the GASDRSE due to its prohibitive computational time besides that it fails at higher modulation orders ($>64$-QAM) as well.  In addition, for QPSK FTN signaling, we also compare the proposed ADMMSE against the SSSSE and the SSSgb\textit{K}SE in \cite{bedeer2017very}, where $K$=3. {Furthermore, ADMMSE is compared against two FDE detectors from \cite{6574905} and \cite{7968313} in QPSK FTN signaling}. The SE is defined here as SE$=\frac{\log_{2}M}{\left(1+\alpha\right)\tau}$, where $M$ is the constellation size \cite{kulhandjian2019low}.  We compare the performance in terms of BER and the practical computational effort by measuring the average CPU time. A comparison of the SE gain for QAM FTN of modulation orders in the range 4-QAM to 64K-QAM is also provided. {The simulations were conducted on an  Intel Core i7-4770 3.4 GHz CPU and 24 GB RAM machine.}

This section is divided into three sub-sections. In Sub-section \ref{QPSK}, we discuss the results for low order modulation, i.e. QPSK. In Sub-section \ref{16-QAM}, moderately high-order 16-QAM results are presented. Finally, in Sub-section \ref{ultra}, the SE gains for the FTN signaling of \textit{M}-QAM modulation orders in the range $64 \leq M\leq 65,536$ using the proposed ADMMSE are provided and discussed. 

\begin{figure}[t]
	\begin{center}
		\centering

		\subfloat[  {$\tau=0.8$}.]{
			\includegraphics[width=0.49\textwidth]{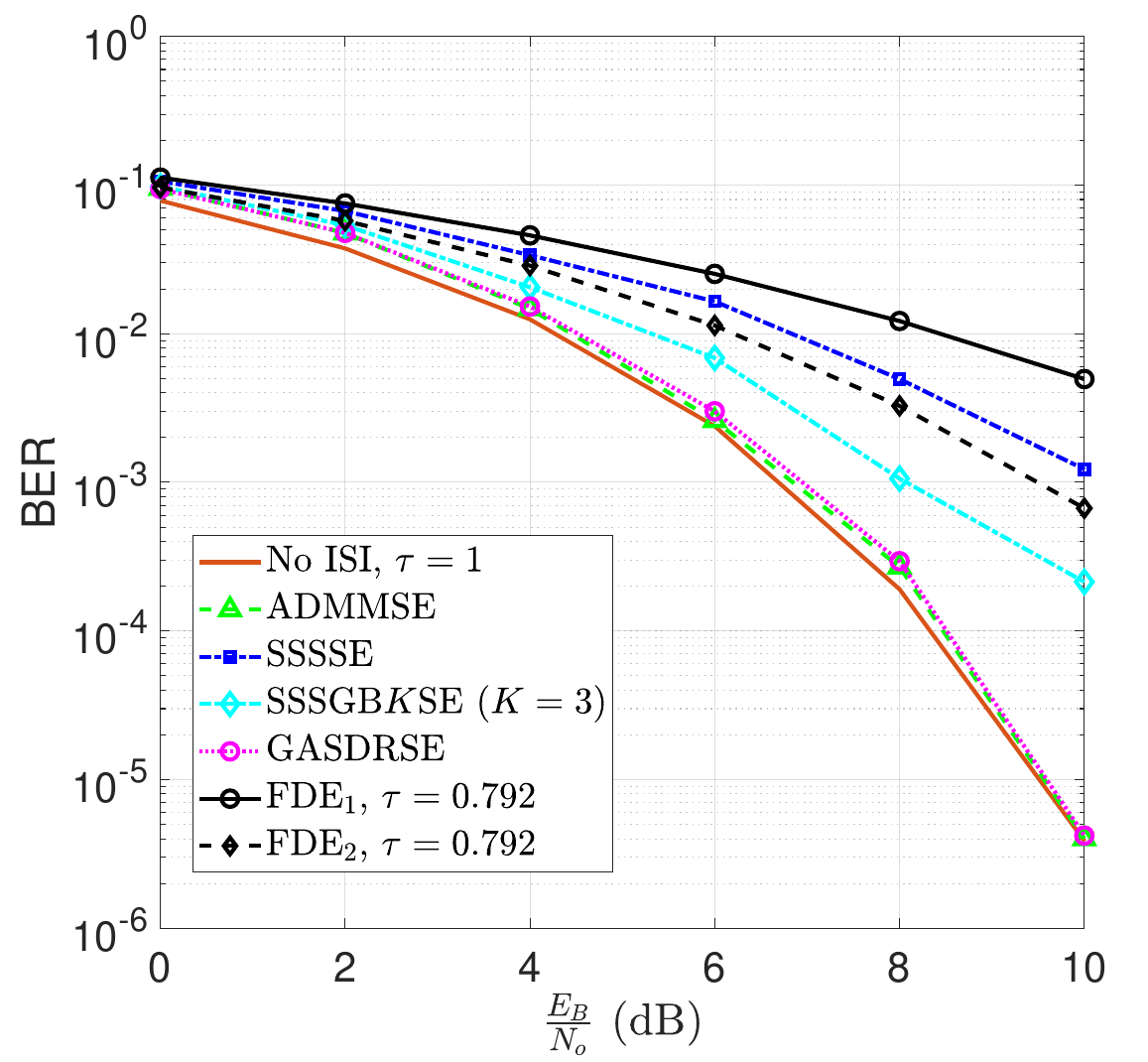}\label{QPSK_0.8}
		}
		\hfill
		\subfloat[$\tau=0.7$.]{
			\includegraphics[width=0.48\textwidth]{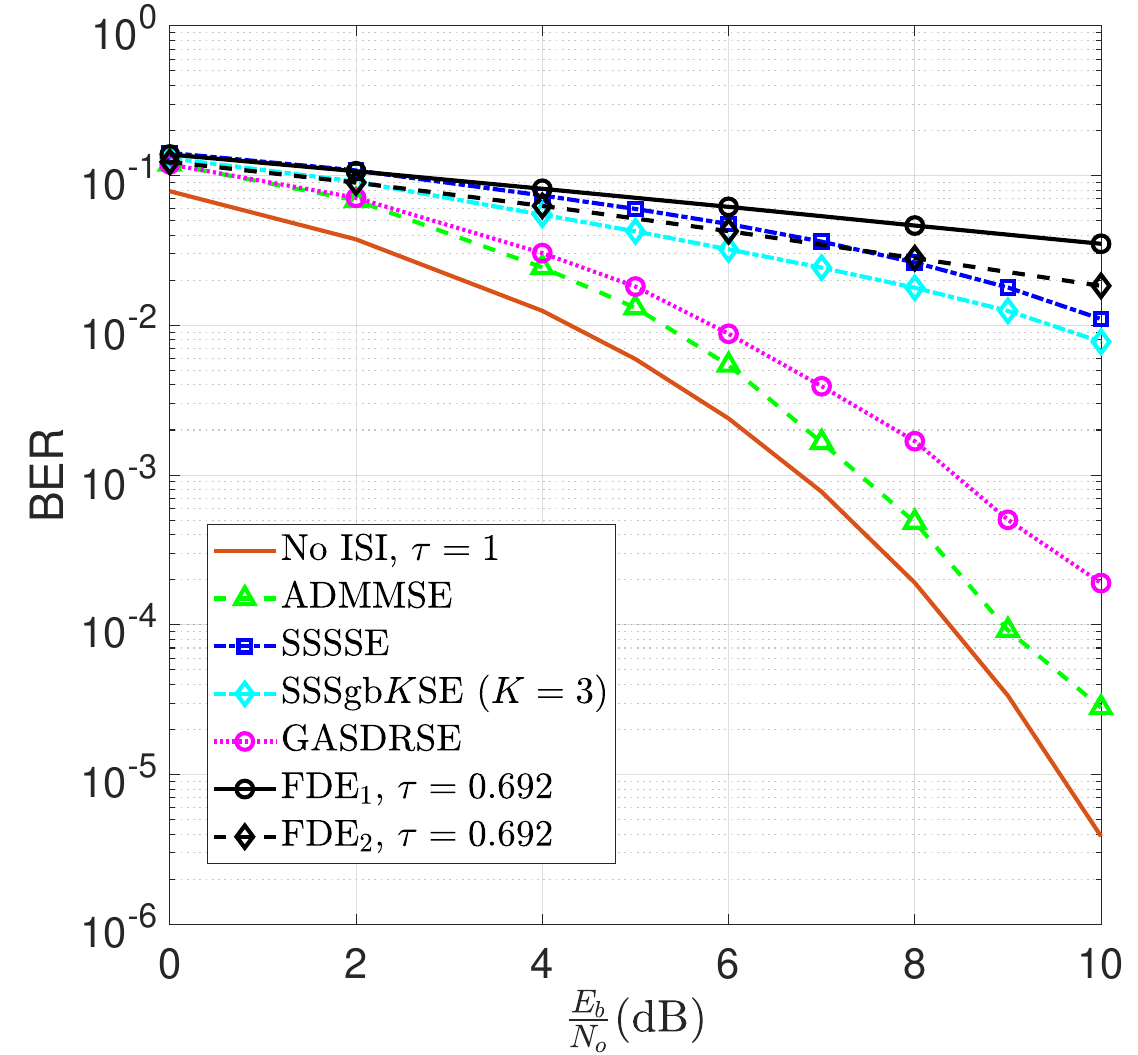}\label{QPSK_0.7}
		}

	\end{center}
	\caption{BER of QPSK FTN signaling detection for $\tau\in\left \{0.7,0.8  \right \}$ for an rRC roll-off factor $\alpha=0.3$ using the proposed ADMMSE versus the SSSSE,  SSSgb\textit{K}SE ($K=3$), GASDRSE, {$\mathrm{FDE_{1}}$ in \cite{6574905}, $\mathrm{FDE_{2}}$ in \cite{7968313}}  and the Nyquist signaling case ($\tau=1$).}
	\label{QPSK_0.7_0.8}
	
\end{figure}

\subsection{QPSK FTN Signaling Detection}\label{QPSK}
\begin{figure}[t]
	\begin{center}
	\centering

     \includegraphics[width=0.49\textwidth]{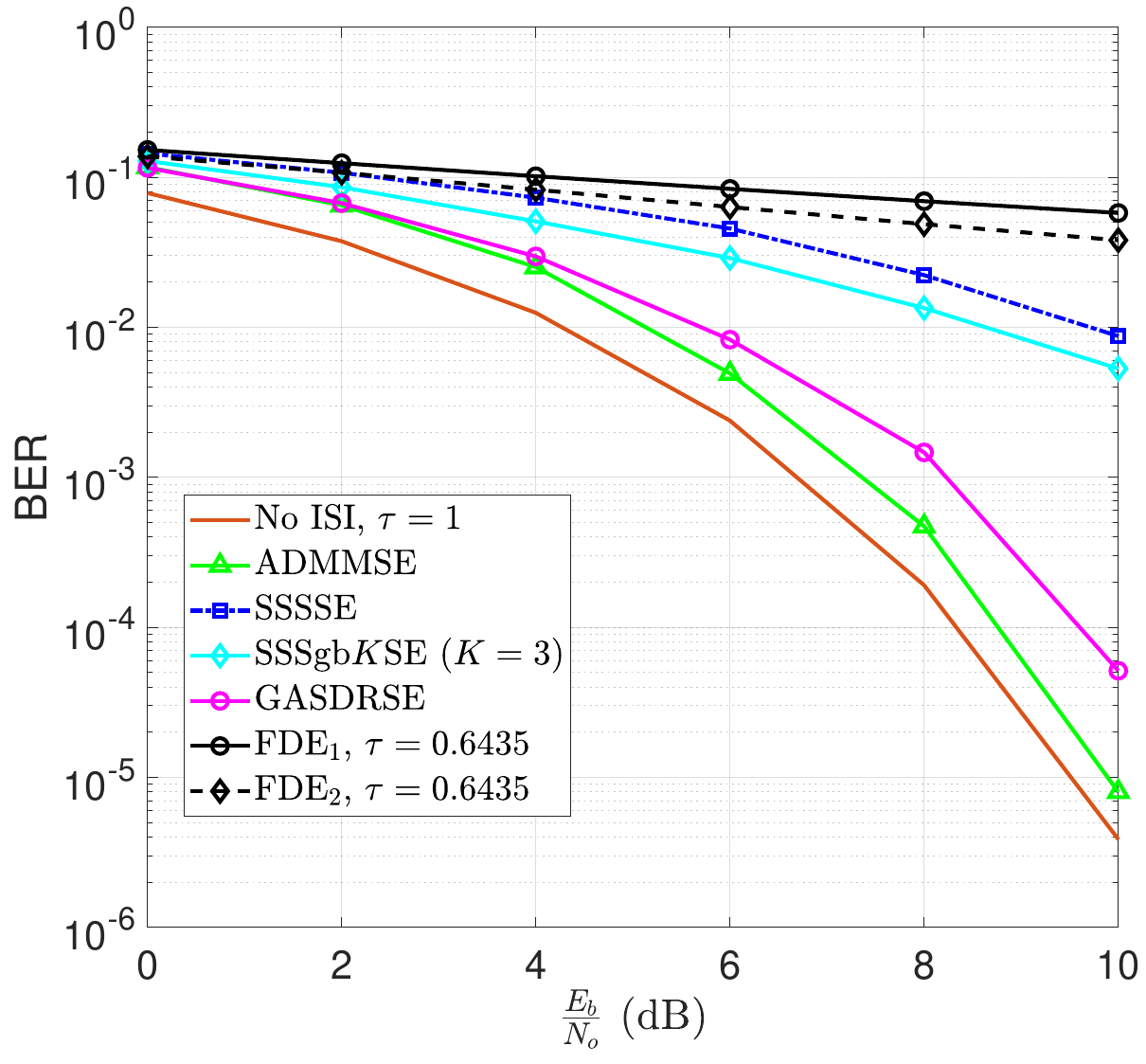}%

\end{center}
\caption{Performance in terms of BER for QPSK FTN signaling detection at $\tau=0.65$ and $\alpha=0.5$ using the proposed ADMMSE scheme versus the SSSSE, SSSgb\textit{K}SE, GASDRSE, {$\mathrm{FDE_{1}}$ in \cite{6574905}, $\mathrm{FDE_{2}}$ in \cite{7968313}} and the Nyquist signaling case ($\tau=1$).}
\label{QPSK_0.65}
\end{figure}

In the simuations conducted for QPSK FTN signaling, the number of ADMM iterations is set to $L=200$,  the number of initializations is $\kappa=50$ and ADMMSE's parameter $\rho=0.5$. Figs. \ref{QPSK_0.7_0.8} and  \ref{QPSK_0.65} depict the BER of QPSK FTN signaling as a function of $\frac{E_b}{N_o}$ for $\tau \in \left \{  0.7, 0.8\right \}$ and $\tau=0.65$ respectively, for the proposed ADMMSE detector versus SSSSE, SSSgb\textit{K}SE ($K=3$), {the GASDRSE, $\mathrm{FDE_{1}}$ in \cite{6574905} and $\mathrm{FDE_{2}}$ in \cite{7968313}} at a rRC roll-off factors of 0.3 and 0.5. We can see that the ADMMSE and GASDRSE perform almost equally at $\tau=0.8$, i.e., at SE$~=1.92$ bits/s/Hz, and outperforming the other detectors. At a smaller acceleration parameter, $\tau=0.7$, the transmitter introduces stronger ISI but with the merit of a higher SE of $2.20$ bits/sec/Hz. Fig. \ref{QPSK_0.7} shows that ADMMSE  outperforms GASDRSE. Fig. \ref{QPSK_0.65} shows the BER for SE$~=2.05$ bits/sec/Hz achieved by  $\tau=0.65$ and $\alpha=0.5$ respectively. It is clear that the performance of ADMMSE is the best.  {It is worth noting that both SSSgb$K$SE and GASDRSE outperform $\mathrm{FDE_{1}}$ in \cite{6574905} and $\mathrm{FDE_{2}}$ in \cite{7968313} by 2.2 dB and 1.2 dB, respectively, for SE=1.92 bits/sec/Hz. We still see that both FDE detectors are inferior to the SSSgb$K$SE and GASDRSE at lower $\tau$. We therefore restrict the rest of the comparisons to the GASDRSE and SSSgb$K$SE where applicable.}

\begin{figure}[t]
	\begin{center}
	\centering
\subfloat[ Linear scale.]{
     \includegraphics[width=0.48\textwidth]{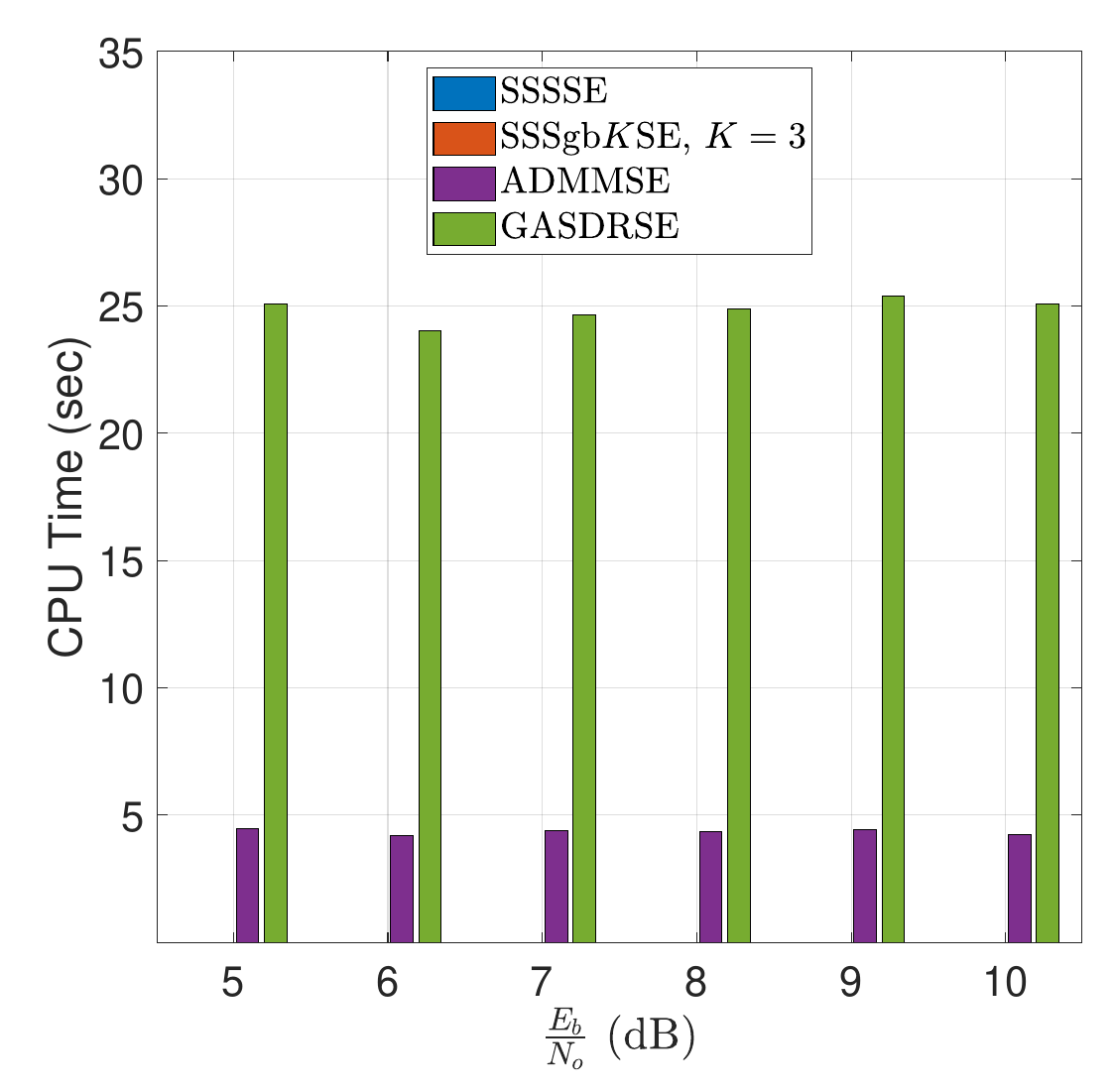}%
    }%
\hfill
\subfloat[ Logarithmic scale.]{
     \includegraphics[width=0.48\textwidth]{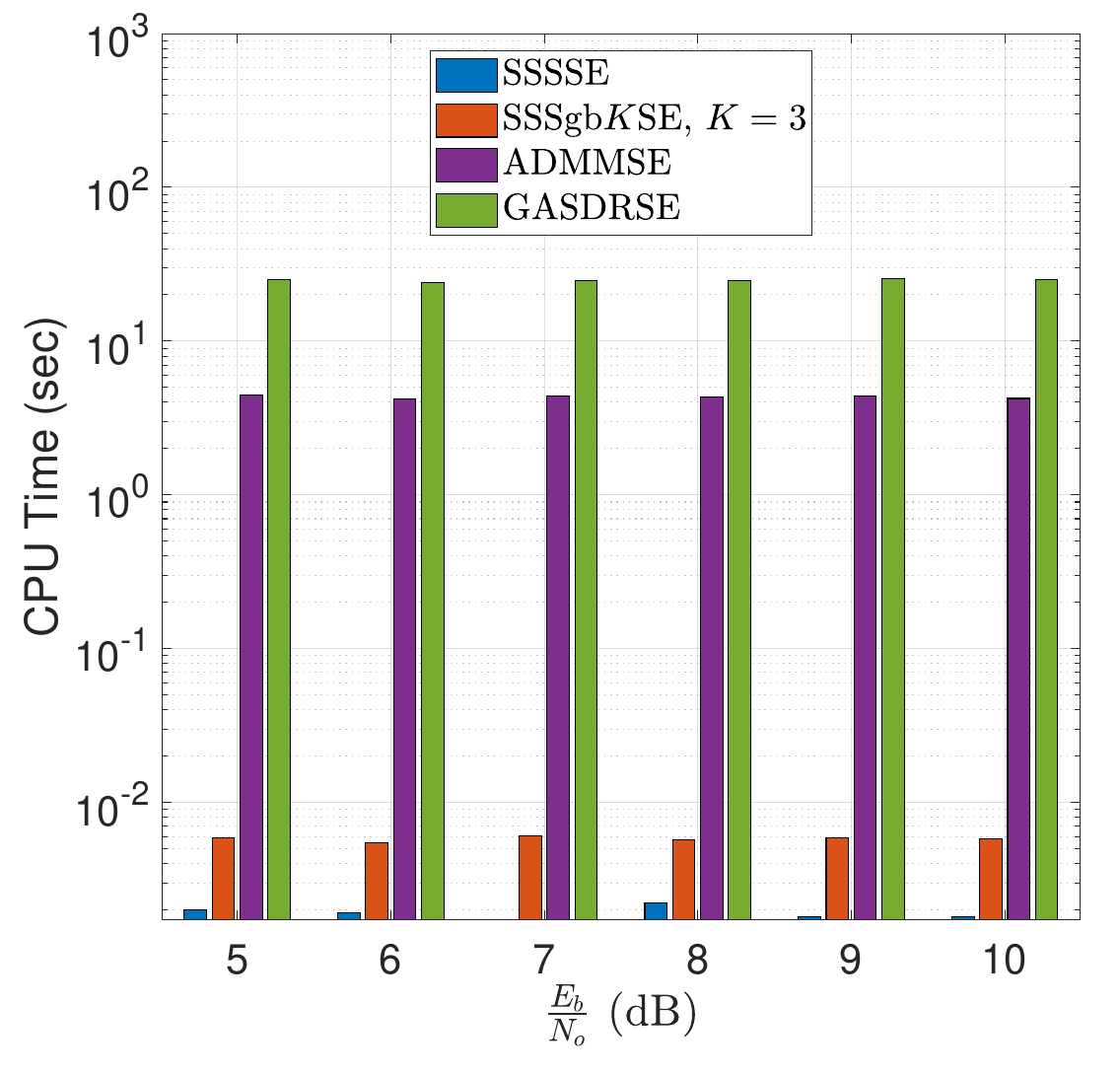}%
    }

\end{center}
\caption{Average CPU time, in seconds, needed by each FTN signaling detector for the detection of QPSK FTN signaling.}
\label{CPU_time_QPSK}
\end{figure}

The computational effort required for the detection of QPSK FTN signaling by the proposed ADMMSE scheme versus the aforementioned benchmark schemes is assessed by gauging the average CPU time. Figure \ref{CPU_time_QPSK} shows that the CPU time of the SSSSE and SSSgb\textit{K}SE ($K=3$) is almost negligible compared to all other algorithms, however  their performance is the poorest. The GASDRSE requires more than 4 times the CPU time required by the ADMMSE, and still the ADMMSE performs better for the values of $\tau$ and $\alpha$ mentioned earlier. 

\begin{figure}[t]
	\begin{center}
	\centering
\includegraphics[width=\columnwidth]{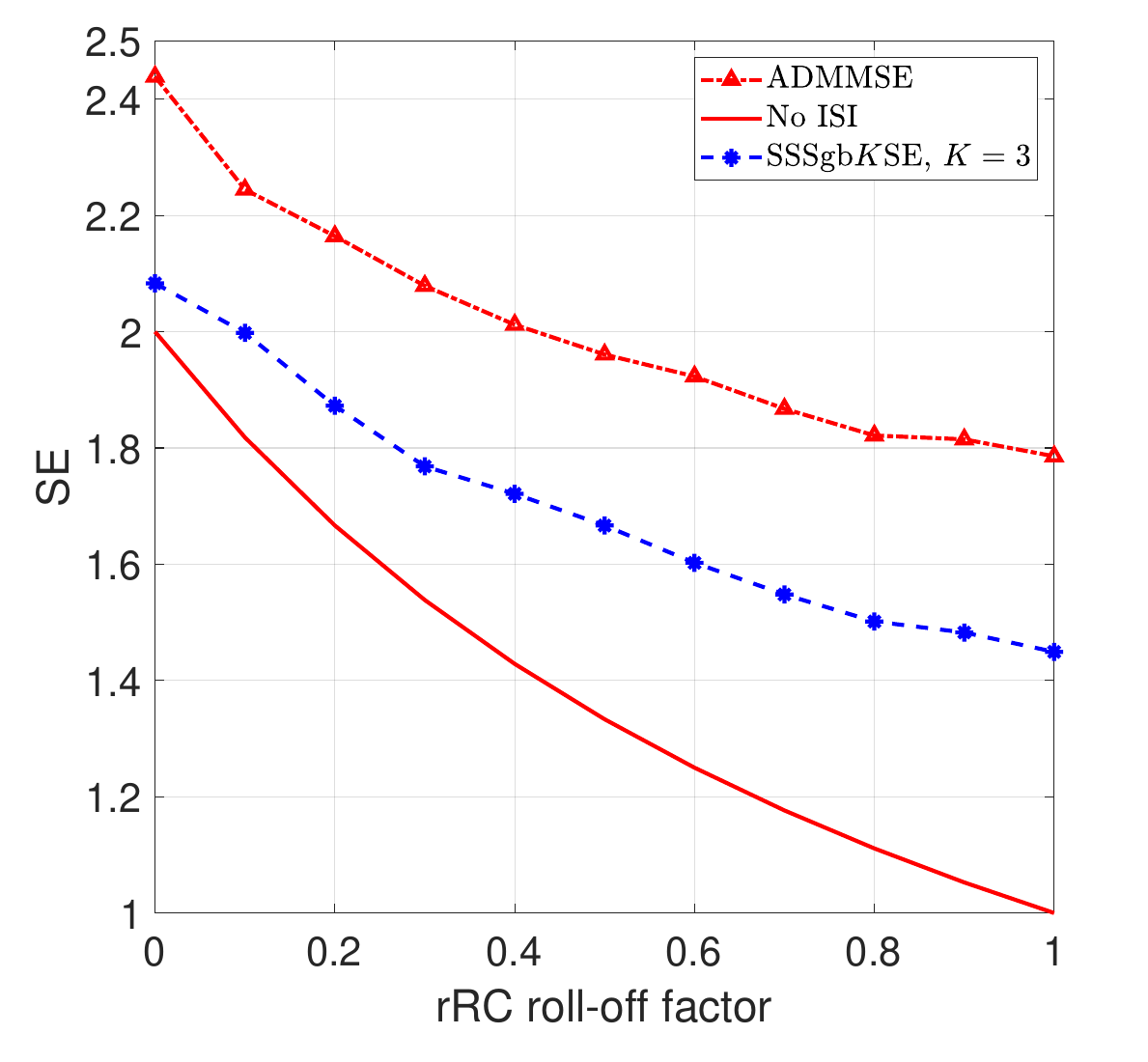}%
\caption{Spectral efficiency comparison of QPSK FTN signaling versus $\alpha$ using the proposed  ADMMSE versus SSSgb\textit{K}SE ($K$=3) and Nyquist signaling at BER $= 10^{-4}$ and the same SNR.}
\label{QPSK_SE}
\end{center}
\end{figure}

In Fig. \ref{QPSK_SE}, we plot the SE of QPSK Nyquist signaling (i.e., no ISI and $\tau =1$) and the proposed ADMMSE for QPSK FTN signaling as a function of the rRC filter roll-off factor $\alpha$. For this simulation, ADMMSE's parameters are set to $\kappa=20$, $L=120$ and $\rho=0.5$. Additionally, we compare the achieved SE using the proposed ADMMSE against that obtained by SSSgb\textit{K}SE ($K=3$). The same SNR is used in all the schemes in the figure that achieves a BER $= 10^{-4}$ for QPSK Nyquist signaling and ADMMSE's parameter is set to $\rho=0.5$. The rRC filter roll-off factor $\alpha$ is varied from 0 to 1 with steps of 0.1. As expected, for all the schemes, the SE decreases as the $\alpha$ increases. For each value of $\alpha$, we conduct an exhaustive  search with a resolution of of $0.01$ for each of the FTN signaling detection schemes in Fig. \ref{QPSK_SE} to find the smallest acceleration parameter $\tau$ at which the BER does not get degraded at the same SNR. In this way, we are able to find the SE that a particular FTN signaling detector could achieve without spending any additional SNR, and without any degradation for the BER in comparison to the Nyquist signaling case. From the figure, one can see that the proposed ADMMSE achieves a spectral efficiency gain of at least $22 \%$ at $\alpha=0$ with respect to the Nyquist case for any value of $\alpha$. We can also see that the ADMMSE FTN signaling detector yields an SE gain over the Nyquist signaling at any roll-off factor with a maximum of $78 \%$ at $\alpha=1$. For $0 \leq \alpha \leq 0.4$, the proposed FTN signaling ADMMSE detector has a higher spectral efficiency in comparison to the Nyquist signaling case. We can also see that ADMMSE achieves a higher spectral efficiency versus SSSgb\textit{K}SE ($K=3$) for all values of $\alpha$ ranging between $17.8 \%$  and $33.64 \%$.

\subsection{High-Order QAM FTN Signaling Detection} \label{16-QAM}

\begin{figure}[t]
	\begin{center}
	\centering
\subfloat[   {$\tau=0.8$}.]{
     \includegraphics[width=0.48\textwidth]{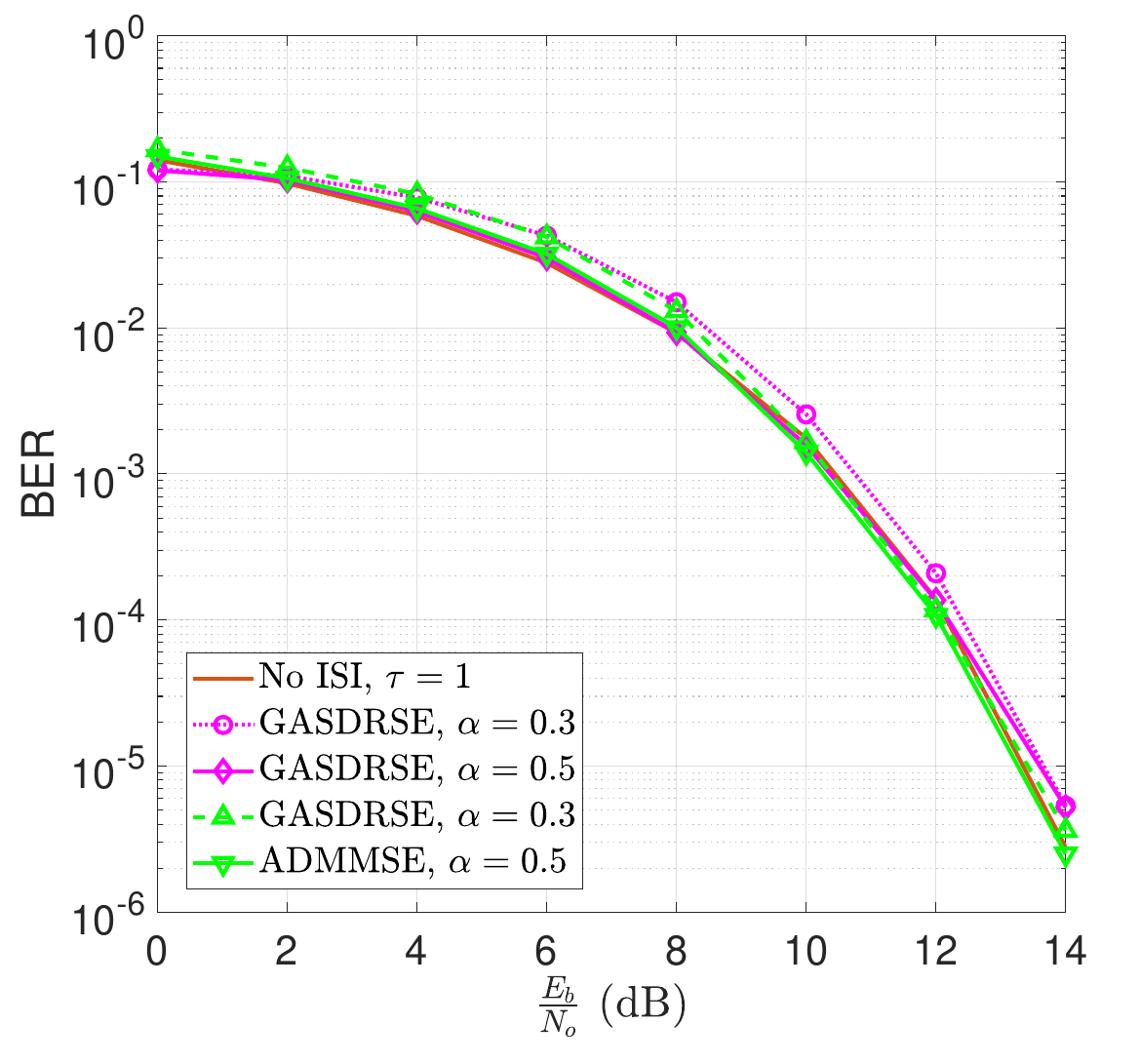}\label{QAM16_0.8}%
    }
\hfill
\subfloat[ $\tau=0.7$.]{
     \includegraphics[width=0.5\textwidth]{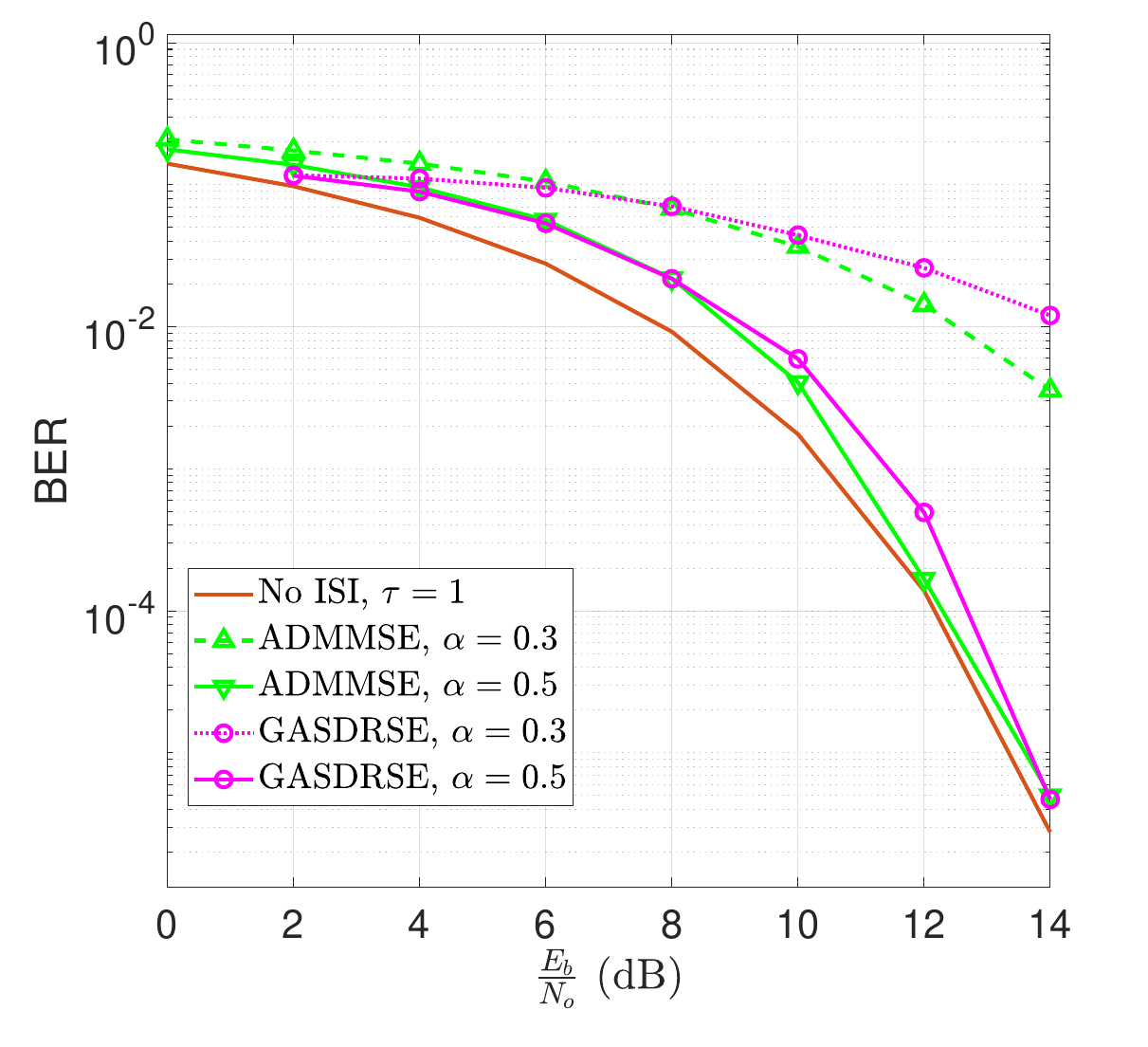}\label{QAM16_0.7}
}%
\end{center}
\caption{ A performance comparison for the 16-QAM FTN signaling detection for $\tau \in \left\{0.7,0.8\right\}$ and rRC roll-off factors $\alpha \in \left\{0.3, 0.5\right\}$ using the proposed ADMMSE versus GASDRSE.}
\label{16-QAM_results}
\end{figure}

In this section, we investigate the performance and practical computational overhead for the detection of high-order QAM modulation FTN signaling using the proposed ADMMSE against GASDRSE. The number of ADMM iterations is set to $L=200$,  the number of initializations $\kappa=50$. Fig. \ref{QAM16_0.8} illustrates the performance in terms of BER for 16-QAM FTN signaling detection for $\tau=0.8$ and $\alpha=\left\{0.5, 0.3\right\}$, i.e., SE$~=3.333$ bits/sec/Hz and SE$~=3.85$ bits/sec/Hz, respectively, where ADMMSE's parameter is $\rho=0.5$. For $\alpha=0.3$, one can see that the ADMMSE achieves a comparable performance slightly better than GASDRSE, for moderate SNR values, and that both converge to the BER of Nyquist signaling case asymptotically. For $\alpha =0.5$ (i.e., for mild ISI), the ADMMSE and GASDRSE achieve approximately the same BER as that of the Nyquist signaling case. It should be noted that the best performance an FTN signaling detection scheme could achieve cannot out perform the Nyquist signaling case in terms of BER. In other words, the lowest BER we can hope for in FTN signaling is equal to that of the Nyquist signaling case at the same SNR.

\begin{figure}[t]
	\begin{center}
	\centering
	{
     \includegraphics[width=0.5\textwidth]{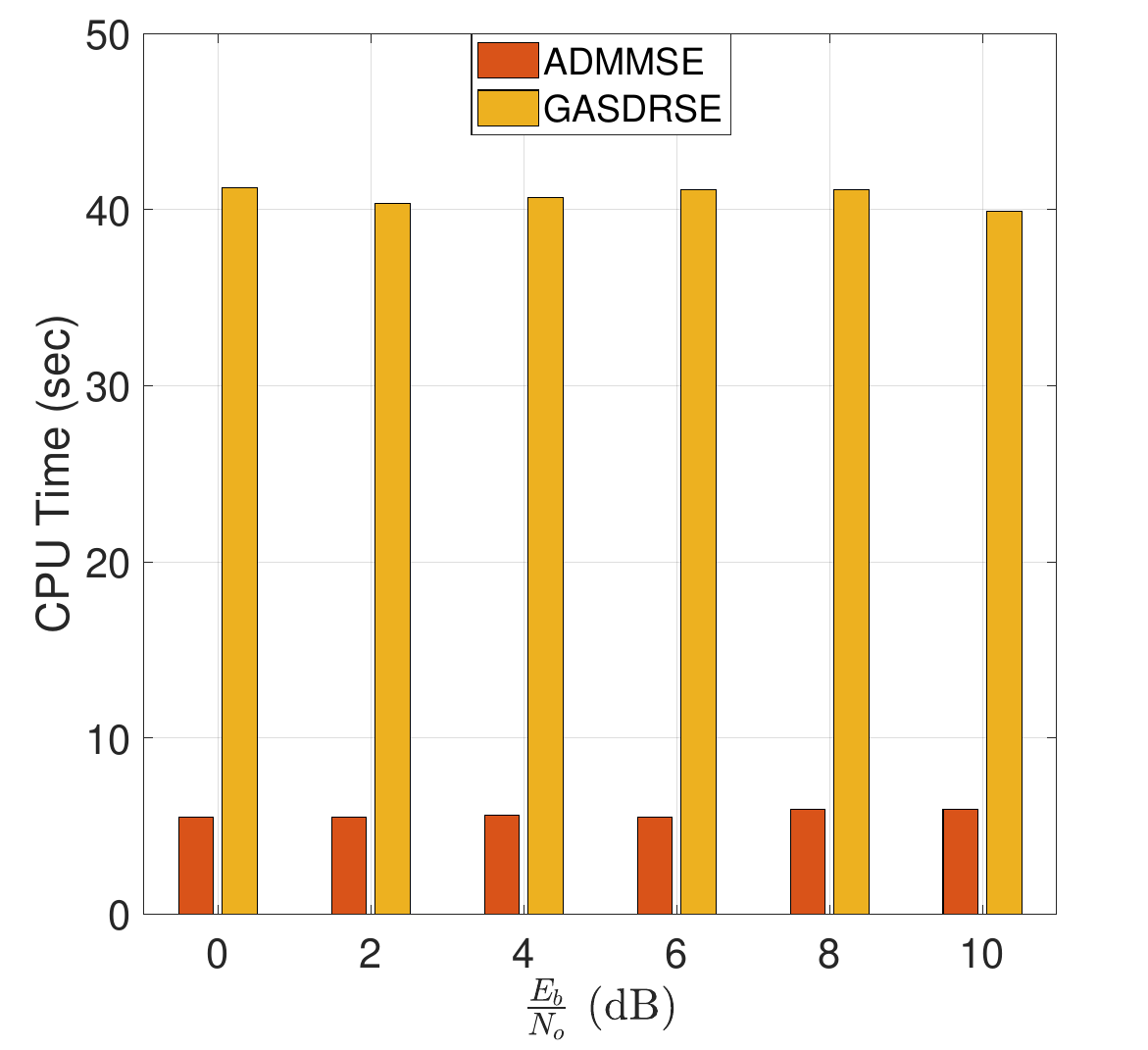} 
    }
\end{center}
\caption{ CPU time of the proposed and benchmark FTN signaling detector for 16-QAM FTN signaling.}
\label{QAM_16_CPU_Time}%
\end{figure}

 In Fig. \ref{QAM16_0.7}, one can see the performance in the case of $\tau=0.7$ and $\alpha=\left\{0.5, 0.3\right\}$,  i.e., SE$~=3.81$ bits/sec/Hz and SE$~=4.4$ bits/sec/Hz, respectively, where ADMMSE's parameter $\rho$ is $0.35$ and $0.2$, respectively. For the case of $\alpha=0.5$, the ADMMSE is only a maximum of 1 dB worse than the Nyquist case at BER of $10^{-2}$, while asymptotically, this gap diminishes. As we can see, the BER of the ADMMSE for $\alpha=0.5$ is slightly better than GASDRSE, which also has a diminshing gap at high SNR with respect to the Nyquist case. For $\tau=0.7$ and $\alpha=0.3$, we see a degradation in performance for ADMMSE and GASDRSE as well, while ADMMSE still performs better than GASDRSE as the SNR increases. The degraded performance experienced when $\tau$ or $\alpha$ are reduced is obviously due to the higher ISI. The CPU time measuring the computational effort of the two detectors is illustrated in Fig. \ref{QAM_16_CPU_Time}. We can clearly see that GASDRSE requires almost $700\%$ extra computational effort in comparison to the ADMMSE.

\begin{figure}[t]
	\begin{center}
	\centering
\includegraphics[width=0.5\textwidth]{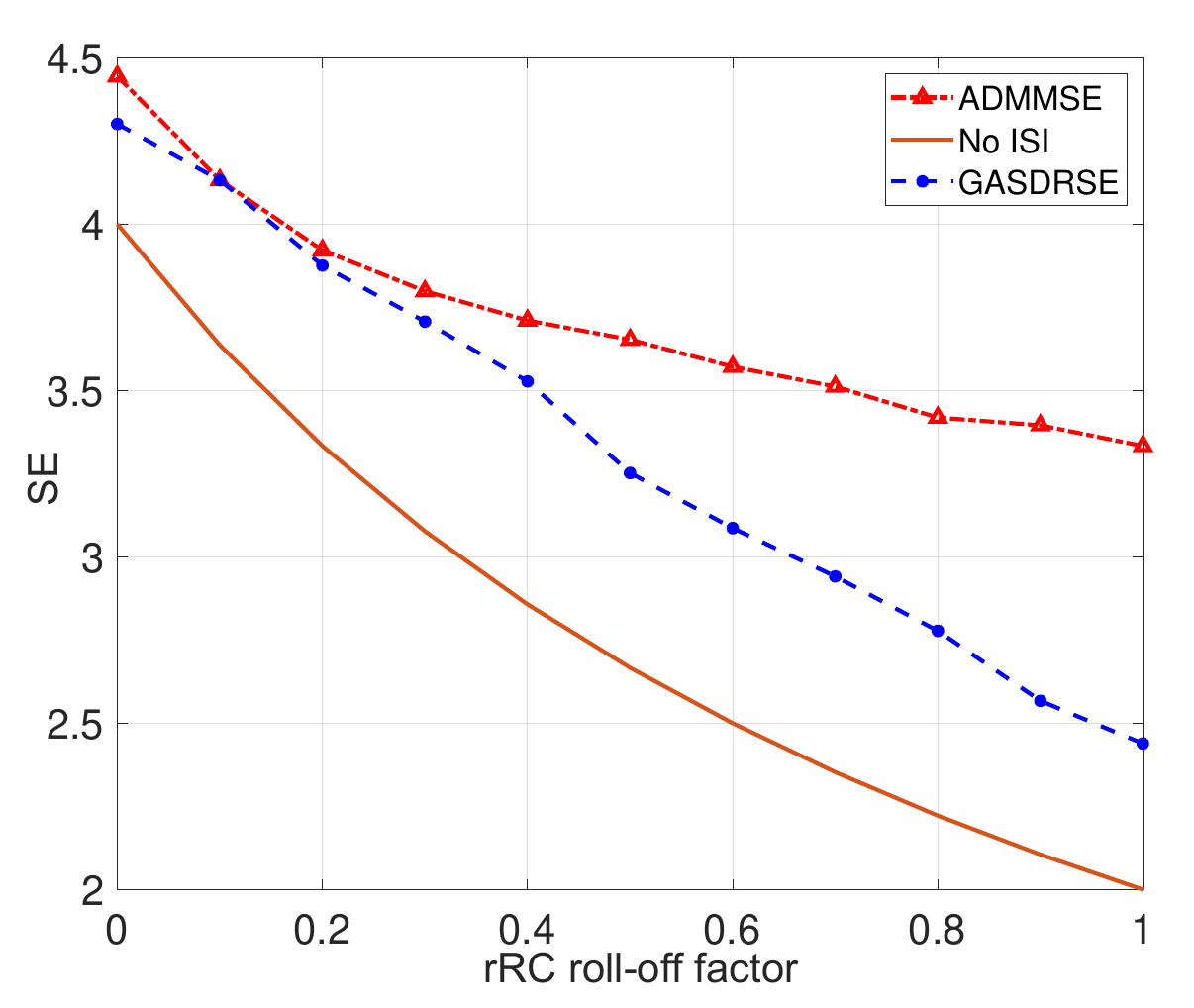}%
\caption{Spectral efficiency comparison of 16-QAM FTN signaling versus $\alpha$ using the proposed  ADMMSE vs GASDRSE and Nyquist signaling at BER $= 10^{-4}$ and same SNR.}
\label{16-QAM_SE}
\end{center}
\end{figure}

In Fig. \ref{16-QAM_SE}, we show the SE for each of the ADMMSE and the GASDRSE FTN signaling detectors for 16-QAM FTN, as well as, the 16-QAM Nyqsuist signaling case. The same SNR and BER$~=10^{-4}$ are considered, and ADMMSE's parameter is set to $\rho=0.5$. Similar to the simulations of the QPSK FTN signaling in Fig \ref{QPSK_SE},  the minimim $\tau$ that would not violate that target BER is achieved by exhaustive search of a resolution of $0.01$ in order to find the achievable SE for both the ADMMSE and GASDRSE for 16-QAM FTN. Again, as expected, the SE in all cases decrease as $\alpha$ increases.  The ADMMSE achieves an SE gain of at least  $11.1\%$ over Nyquist signaling no matter what values of $\alpha$ the Nyquist signaling would operate at. This SE gain, with respect to the Nyquist signaling case, reaches $66.7 \%$ at $\alpha=1$, while at $\alpha=0.3$ the SE gain is $23.5 \%$. We can also see that at any given $\alpha$, the SE for ADMMSE in 16-QAM FTN is higher than the corresponding SE of Nyquist signaling and that SE improvement increases as $\alpha$ increases, where the maximum improvement is for $\alpha=1$. In comparison to GASDRSE, we can see that the gains of ADMMSE are slightly better for very low $\alpha$ but ADMMSE becomes much superior for moderate to high $\alpha$ as the figure shows, beating GASDRSE by an extra gain of $44.7 \%$ at $\alpha =1$. This is achieved at the much lower CPU overhead as seen in Fig. \ref{QAM_16_CPU_Time}.

\subsection{Ultra High-Order QAM FTN Signaling Detection}\label{ultra}
  {In this section, we demonstrate the BER and SE performance, as well as, the CPU running time of the proposed ADMMSE FTN signaling detection for 256, 1024, 4096, 16,384 (16K), and 65,536 (64K)-QAM.} To the best of our knowledge, the results presented in this section are the first in the FTN signaling literature of their kind, thanks to the very low complexity of ADMMSE whose computational effort has little sensitivity to the modulation order. 

\begin{figure}[t]
	\begin{center}
	\centering
\subfloat[ BER v. $\frac{E_b}{N_o}$ of the ADMMSE.]{
     \includegraphics[width=0.48\textwidth]{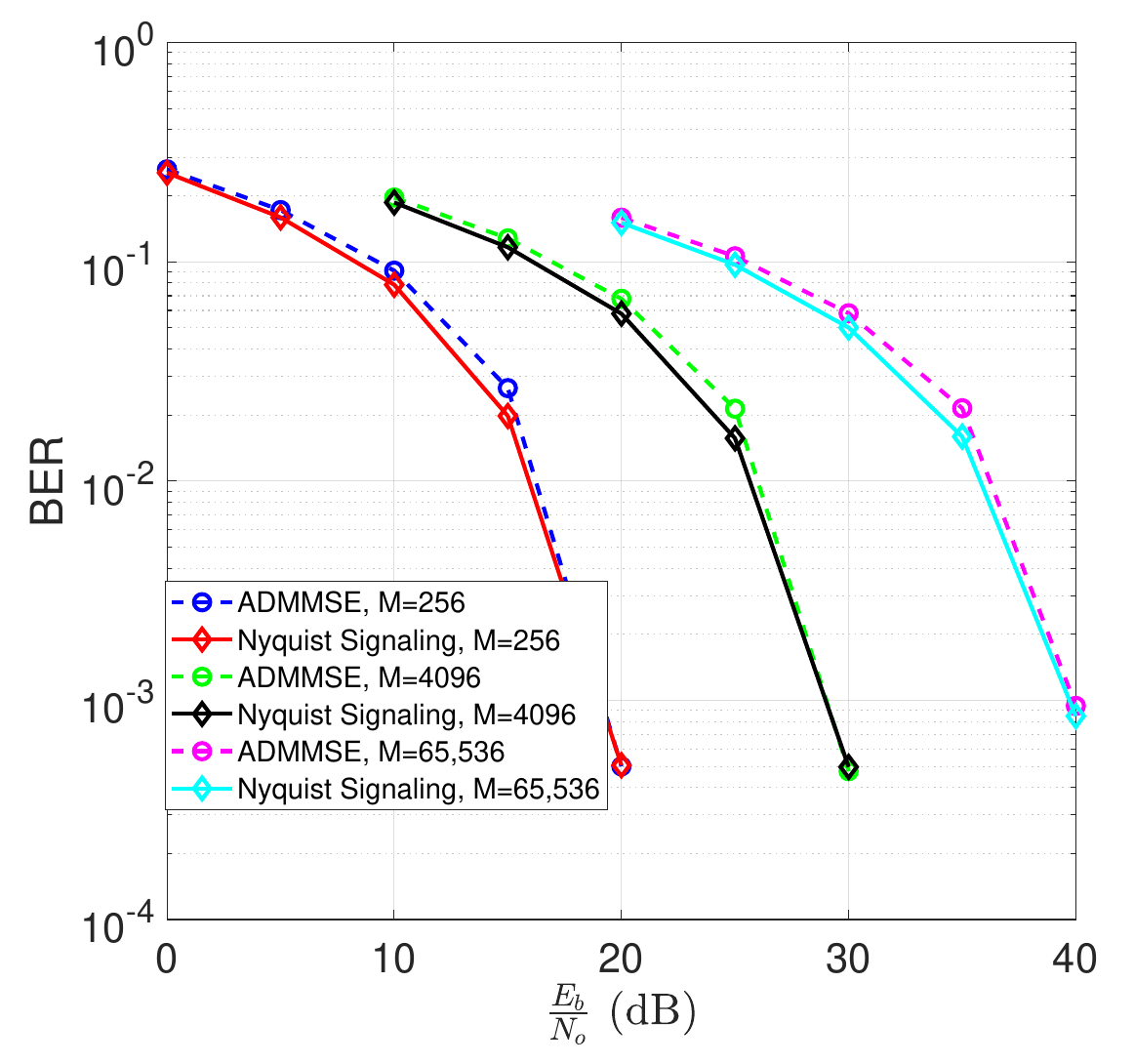}\label{BER_Ultra}%
    }
\hfill
\subfloat[  CPU time of the ADMMSE.]{
     \includegraphics[width=0.48\textwidth]{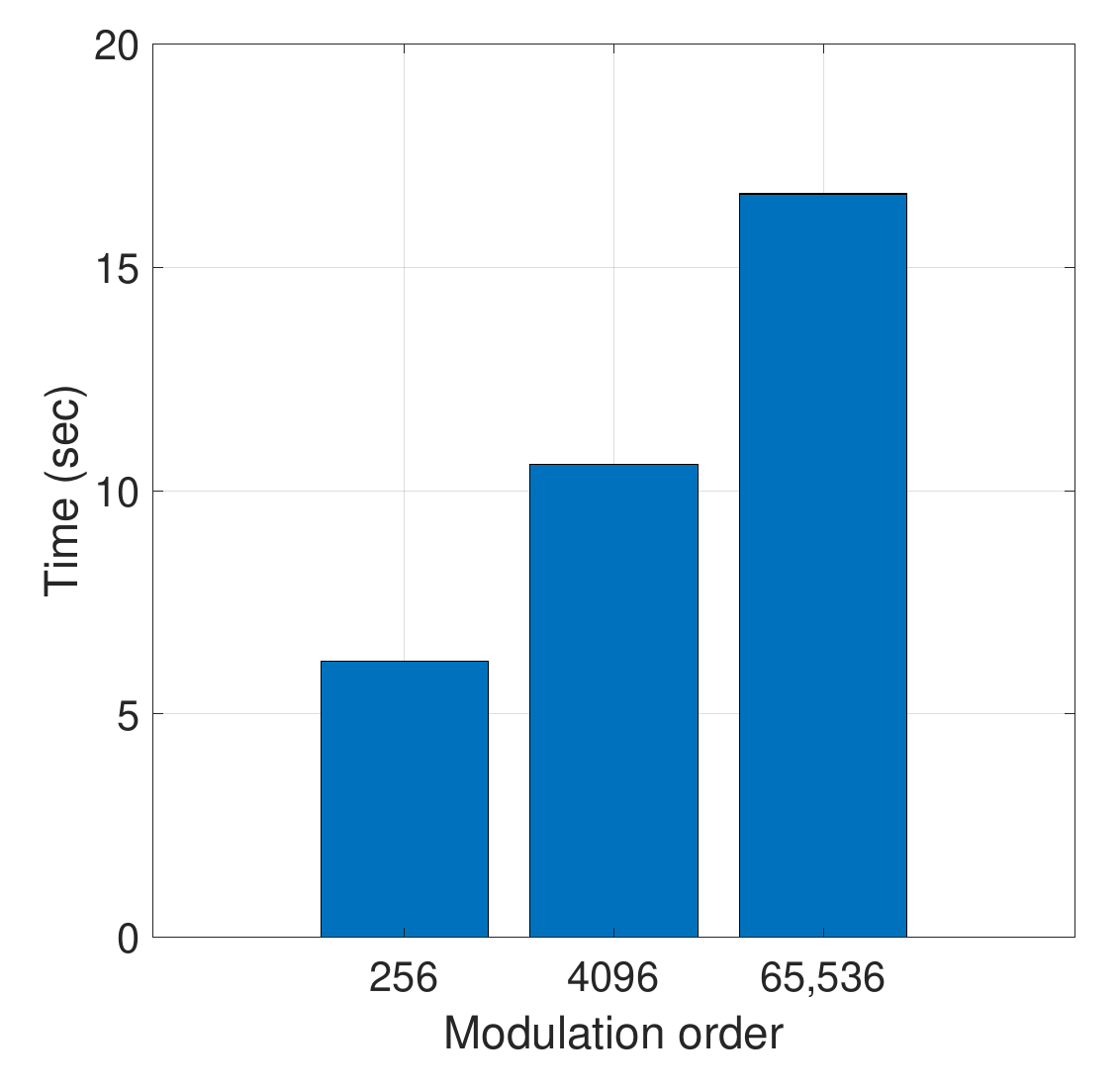}\label{CPU_Ultra}
}%
\end{center}
\caption{  {The performance and computational effort for the 256-QAM, 4096-QAM and 65,536-QAM FTN signaling detection for $\tau= 0.85$ and rRC roll-off factor $\alpha =0.3$ using the proposed ADMMSE.}}
\end{figure}

  {The BER versus $\frac{E_b}{N_o}$ as well as the average CPU time for all the SNRs for the modulation orders 256, 4096 and 65,536 are illustrated in Figs. \ref{BER_Ultra} and \ref{CPU_Ultra}, respectively, for $\tau=0.85$ and $\alpha=0.3$. We can see that for high order modulations the proposed ADMMSE FTN signaling detector does not fail and that as the SNR increases the performance is almost the same as the corresponding Nyquist signaling modulations. Additionally, the CPU running time results  show the low sensitivity of the computational effort with respect to the modulation orders.}
\begin{figure*}[t]
	\begin{center}
	\centering
\subfloat[SE for 64-QAM and 256-QAM.]{
     \includegraphics[width=0.45\textwidth]{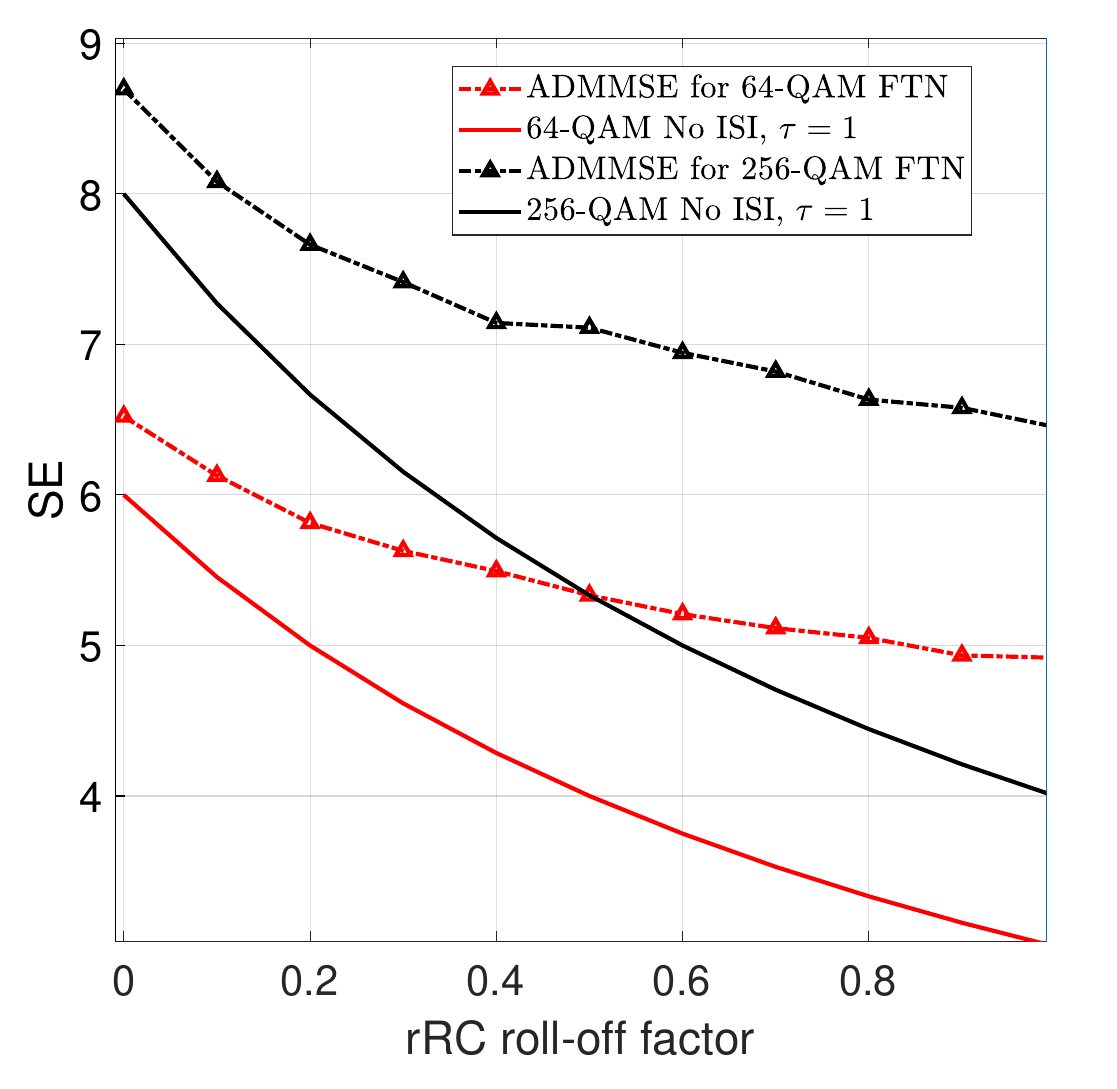}\label{256}
}%
\hfill
\subfloat[SE for 1024-QAM and 4096-QAM.]{
     \includegraphics[width=0.45\textwidth]{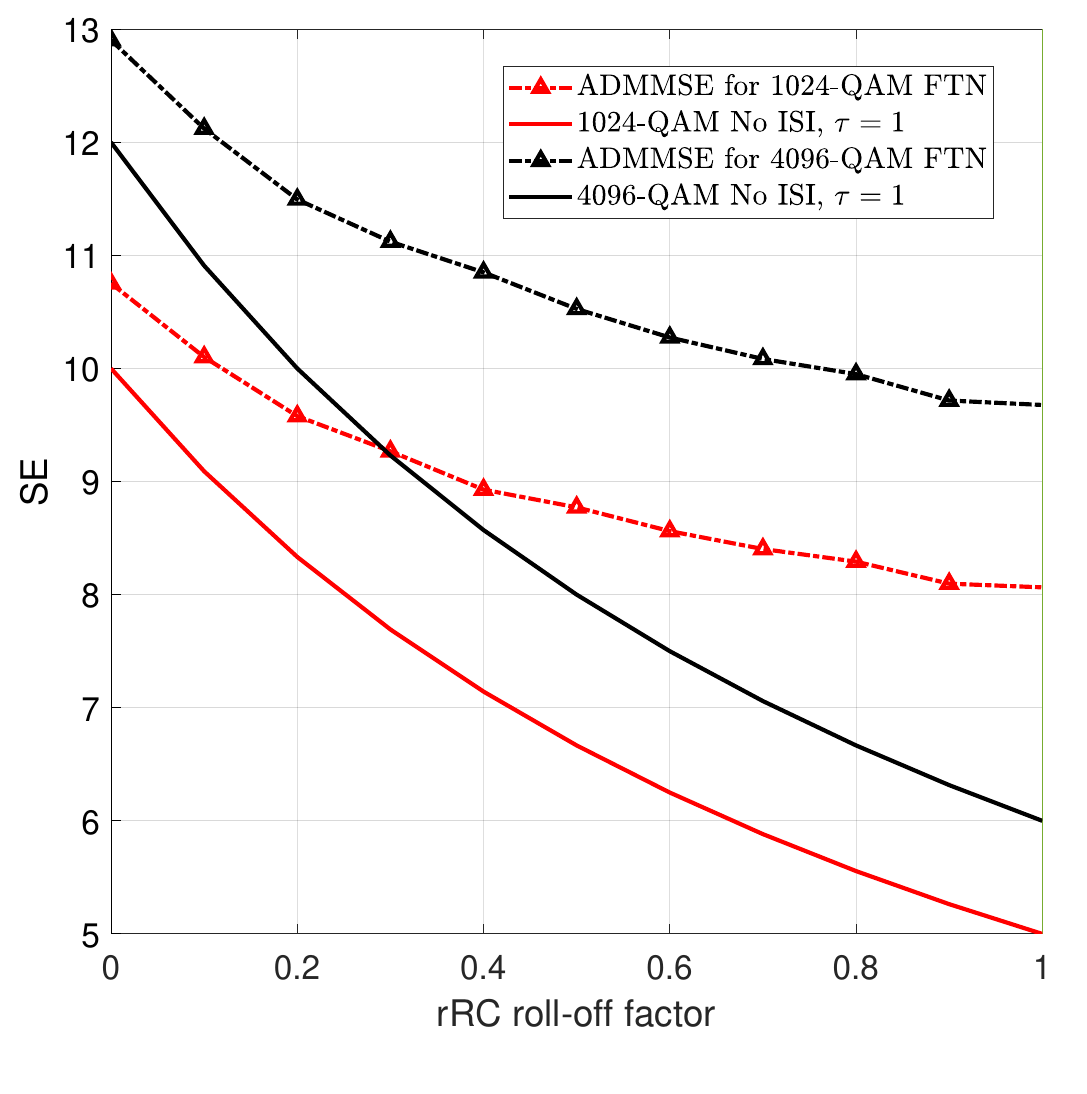}\label{4096}%
    }
\quad
\subfloat[SE for 16K-QAM and 64K-QAM.]{
     \includegraphics[width=0.45\textwidth]{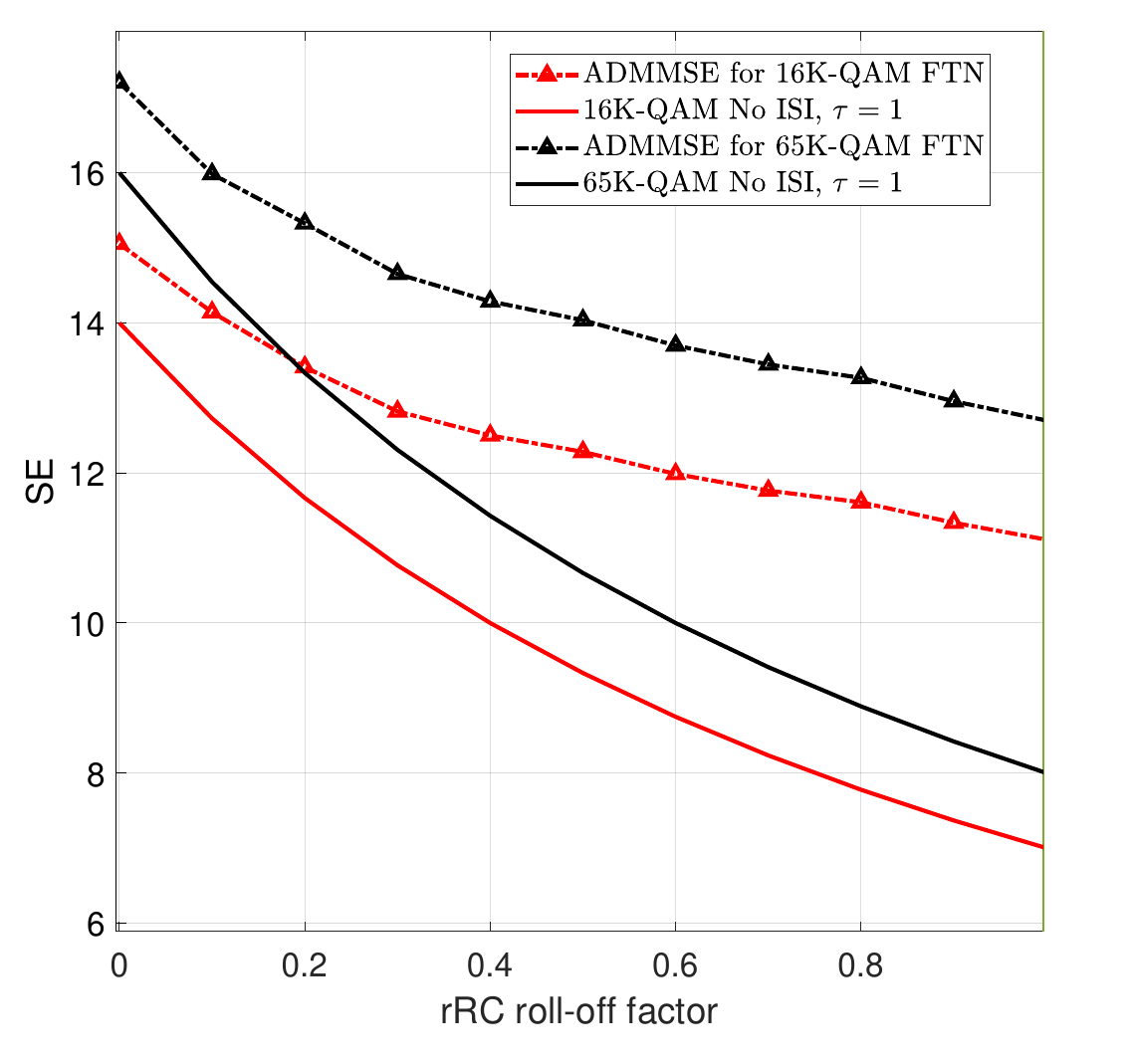}\label{64K}%
    }
\hfill
\subfloat[ SE gain (\%) for QPSK FTN signaling up to 64K-QAM FTN signaling versus the rRC roll-off factor ($\alpha$) achieved using the proposed ADMMSE.]{
     \includegraphics[width=0.45\textwidth]{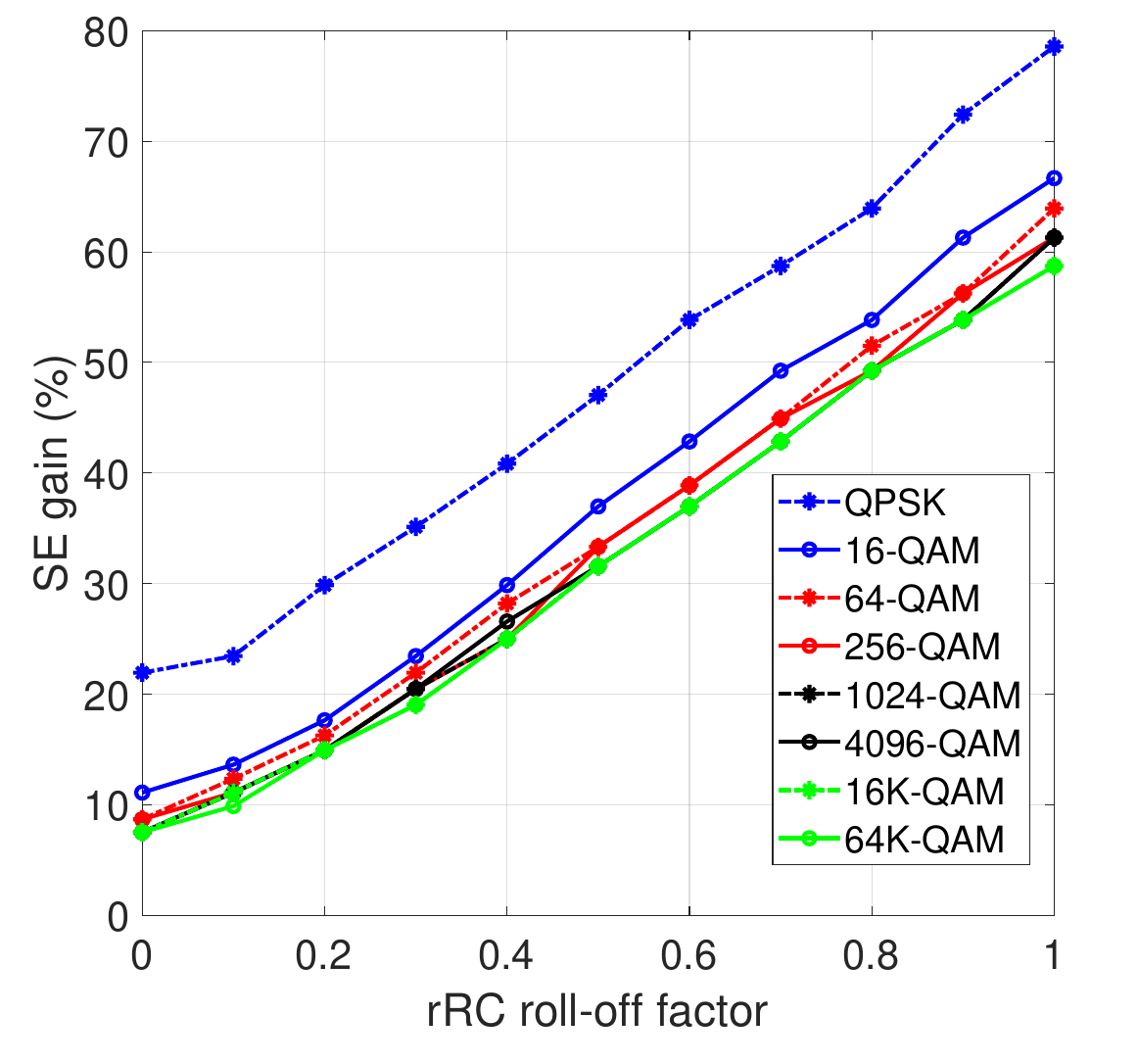}\label{SE_percent}%
    }
\end{center}
\caption{The SE and SE gain for FTN signaling using the ADMMSE detector with respect to Nyquist signaling, for QAM modulation orders in the range $4 \leq M \leq 65,536$, at BER $= 10^{-4}$ and at same SNR values.}
\label{SE_Results}
\end{figure*}

Fig. \ref{256} shows the SEs  for $M=64$ and $M=256$; Fig. \ref{4096} shows the SEs for $M=1024$ and $M=4096$; Fig. \ref{64K} show SEs for $M=16,384$ and $M=65,536$; while Fig. \ref{SE_percent} shows the SE gains in (\%) for $M =4$ up to $M=65,536$. For each modulation order, the same SNR value is used for both the FTN signaling with the ADMMSE detector and the Nyquist signaling cases that achieves a BER of $10^{-4}$.  For the ADMMSE, the minimum $\tau$ at each value of $\alpha$ that maximizes the FTN signaling SE without degrading the BER is found by brute force search. We can see that, as expected, the SE of both the FTN signaling using ADMMSE detector and the Nyquist signaling case decreases as $\alpha$ increases, but the SE gain in \% of the FTN signaling using the ADMMSE detector with respect to the Nyquist signaling case increases linearly as Fig. \ref{SE_percent} shows. Fig. \ref{SE_Results} shows an operating region $0\leq\alpha\leq0.1$ for $M\geq64$ within which the proposed scheme outperforms the SE of the Nyquist signaling case if it were to use any rRC roll-off factor. In Fig. \ref{SE_percent}, we can see the the SE gains achieved by our proposed ADMMSE are the highest for QPSK, and decrease as $M$ increases. For a modulation order of $M= 65536$, the gain varies between $7.5 \%$ at $\alpha=0$ and $58 \%$ for $\alpha =1$, with SE gain of $19.1 \%$ for $\alpha =0.3$. This is approximately an equal gain to what can be achieved by the GASDRSE for only 16-QAM, at a much lower complexity.

Based on these results, we conclude that ADMMSE is suggested over GASDRSE due to its higher SE gains and for its practically lower CPU time that enables its implementation for QAM modulation of orders much beyond $M=16$ (ultra high-order) up to $M=65,536~(64\text{K})$.

\begin{figure}[t]
	\begin{center}
	\centering
\includegraphics[width=0.5\textwidth]{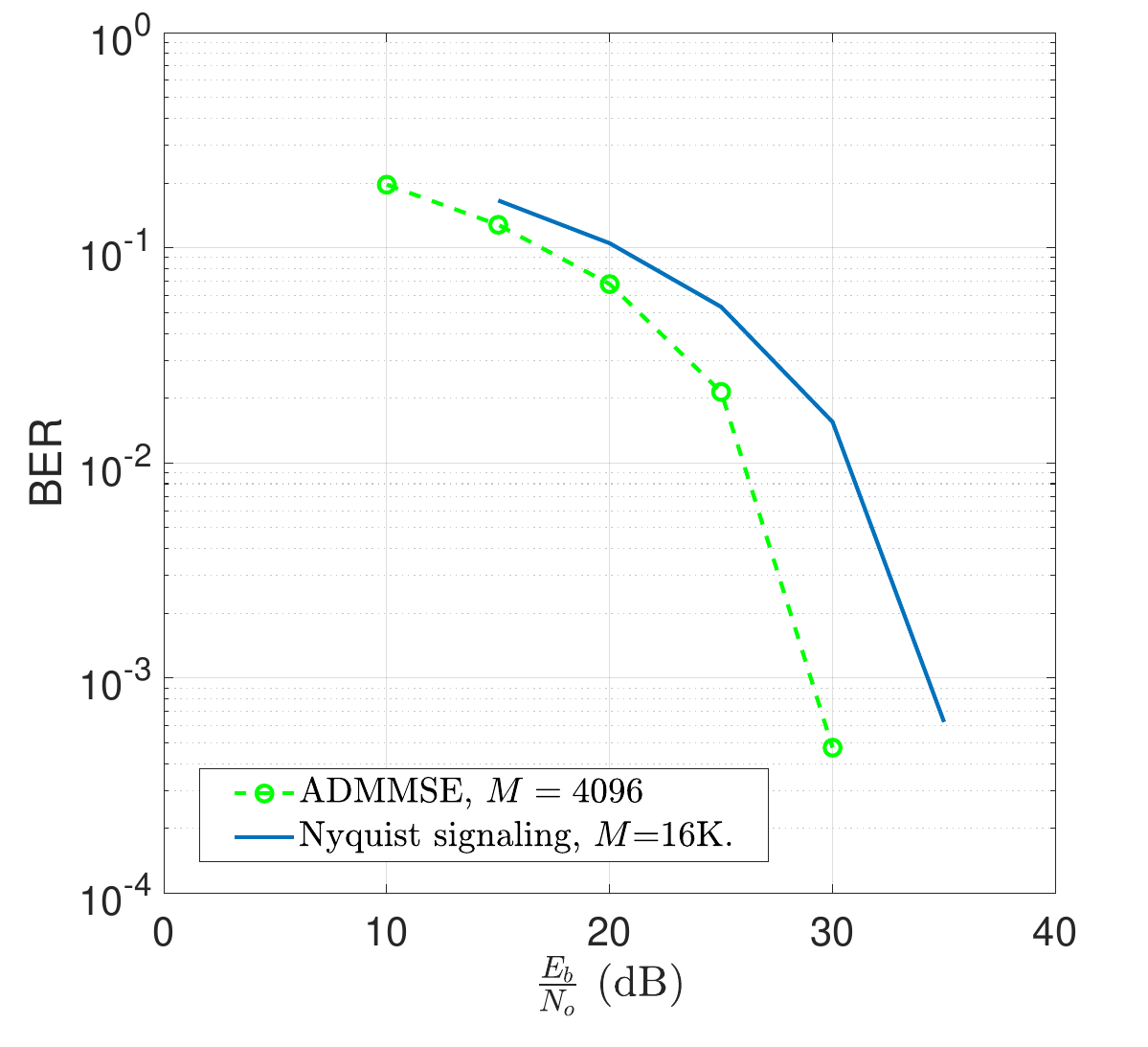}%
\caption{{BER comparison of 4096-QAM FTN signaling with $\tau = 0.857$, using the proposed ADMMSE detector, versus 16K-QAM  Nyquist signaling at SE=10.76 bits/sec/Hz. }}
\label{FTN_vs_Nyquist}
\end{center}
\end{figure}

{To further show the merits of the proposed ADMMSE to detect FTN signaling, we compare in Fig. \ref{FTN_vs_Nyquist} the BER performance of the conventional Nyquist transmission and the proposed ADMMSE detector of ultra high-order FTN while both operate at the same SE. In particular,  we consider 16,384-QAM Nyquist signaling  with roll-off factor $\alpha=0.3$ that operates at a SE value of $\frac{\log_{2}M}{\left(1+\alpha\right)} = 10.76$ bits/sec/Hz. To reach the same SE value of a 4096-QAM FTN signaling, the acceleration factor should be 0.857, i.e., $SE=\frac{\log_{2}M}{\tau\left(1+\alpha\right)} = 10.76$. As can be seen, FTN can complement the performance of high-order modulation and reach the same SE values of Nyquist transmission with much less transmit power.}

\section{Conclusion}\label{Conclude}
FTN signaling is a promising technology that improves SE in next generation communication systems. In this paper, we  proposed a novel FTN signaling detection algorithm that exploits a variant of the ADMM algorithm from the field of convex optimization, that makes it a very suitable technique for the non-convex FTN signaling detection problem. The ADMMSE is proposed in this paper to detect ultra high-order QAM FTN signaling.  The proposed detector achieves excellent performance at a very low complexity, enabling it to obtain excellent SE gains for ultra high-order QAM modulation orders reaching up to 65,536 while being suitable for practical implementation in terms of computational overhead. The ADMMSE demonstrated superiority in QPSK FTN signaling detection over the GASDRSE, the SSSSE and the SSSgb\textit{K}SE in terms of BER at high ISI (i.e., higher spectral efficiency). It also demonstrated a much lower computational effort, measured in CPU time, in comparison to GASDRSE but higher than that of SSSSE and SSSgb\textit{K}SE.
Additionally, we showed that in comparison to GASDRSE for 16-QAM FTN signaling, ADMMSE returns a slightly better performance at a much lower computational effort by $700 \%$, when $\tau \in \left\{0.8, 0.7 \right\}$ and $\alpha \in \left\{0.5, 0.3\right\}$.  The SE of the ADMMSE in comparison to GASDRSE was evaluated for rRC roll-off factors in the range 0 to 1 in 16-QAM FTN signaling showing the superiority of ADMMSE at all roll-off factors, especially moderate to high values.  

Most importantly, we believe that this paper is the first in the FTN signaling literature to propose a detector that is successful in detecting ultra high-order QAM FTN signaling for modulation orders up to 65,536 (64K) at a very low computational complexity. SE gains ranging between $7.5\%$ and $58\%$ were achieved for 64K-QAM FTN signaling. The computational time needed in the detection of  64K-QAM is even much lower than that needed by GASDRSE for 16-QAM. Thanks to the little significance of the ADMMSE's computational effort towards the modulation order. 

{In future work, we plan to extend the proposed ADMMSE to frequency-selective fading channels and investigate its performance in the presence of channel coding, as well as practical impairments such as synchronization errors.}
\section{Acknowledgments}
This work was sponsored by the Corporate University Donations program of the Nokia-Bell Labs. The authors would like to thank Professor Haris Gacanin, Director of Chair for Distributed Signal
Processing, RWTH Aachen University, Germany.

\bibliographystyle{IEEEtran}
\bibliography{IEEEabrv,Accepted_final_version}

\begin{thebibliography}{10}
\providecommand{\url}[1]{#1}
\csname url@samestyle\endcsname
\providecommand{\newblock}{\relax}
\providecommand{\bibinfo}[2]{#2}
\providecommand{\BIBentrySTDinterwordspacing}{\spaceskip=0pt\relax}
\providecommand{\BIBentryALTinterwordstretchfactor}{4}
\providecommand{\BIBentryALTinterwordspacing}{\spaceskip=\fontdimen2\font plus
\BIBentryALTinterwordstretchfactor\fontdimen3\font minus
  \fontdimen4\font\relax}
\providecommand{\BIBforeignlanguage}[2]{{%
\expandafter\ifx\csname l@#1\endcsname\relax
\typeout{** WARNING: IEEEtran.bst: No hyphenation pattern has been}%
\typeout{** loaded for the language `#1'. Using the pattern for}%
\typeout{** the default language instead.}%
\else
\language=\csname l@#1\endcsname
\fi
#2}}
\providecommand{\BIBdecl}{\relax}
\BIBdecl

\bibitem{ourPat}
A.~Ibrahim, E.~Bedeer, and H.~Yanikomeroglu, ``Detector for faster than
  {Nyquist} transmitted data symbols,'' U.S. Provisional Patent 63/150,798,
  Feb. 18, 2021.

\bibitem{itu2017minimum}
ITU-R, ``{Minimum requirements related to technical performance for IMT-2020
  radio interface(s)},'' Nov. 2017.

\bibitem{itu2020}
ITU-T, ``{Focus group on technologies for network 2030: Representative use
  cases and key network requirements},'' Feb. 2020.

\bibitem{8970173}
A.~{Clemm}, M.~T. {Vega}, H.~K. {Ravuri}, T.~{Wauters}, and F.~D. {Turck},
  ``Toward truly immersive holographic-type communication: Challenges and
  solutions,'' \emph{IEEE Communications Magazine}, vol.~58, no.~1, pp. 93--99,
  Jan. 2020.

\bibitem{8387210}
C.~{Han} and Y.~{Chen}, ``Propagation modeling for wireless communications in
  the terahertz band,'' \emph{IEEE Communications Magazine}, vol.~56, no.~6,
  pp. 96--101, Jun. 2018.

\bibitem{8387211}
I.~F. {Akyildiz}, C.~{Han}, and S.~{Nie}, ``Combating the distance problem in
  the millimeter wave and terahertz frequency bands,'' \emph{IEEE
  Communications Magazine}, vol.~56, no.~6, pp. 102--108, Jun. 2018.

\bibitem{anderson2013faster}
J.~B. Anderson, F.~Rusek, and V.~{\"O}wall, ``Faster-than-{N}yquist
  signaling,'' \emph{Proceedings of the IEEE}, vol. 101, no.~8, pp. 1817--1830,
  Mar. 2013.

\bibitem{nyquist1928certain}
H.~Nyquist, ``Certain topics in telegraph transmission theory,''
  \emph{Transactions of the American Institute of Electrical Engineers},
  vol.~47, no.~2, pp. 617--644, Apr. 1928.

\bibitem{mazo1975faster}
J.~E. Mazo, ``Faster-than-nyquist signaling,'' \emph{The Bell System Technical
  Journal}, vol.~54, no.~8, pp. 1451--1462, Oct. 1975.

\bibitem{RACOM}
\BIBentryALTinterwordspacing
RACOM. Racom products: Ray-microwave link. Last accessed 10-11-2020. [Online].
  Available: \url{https://www.racom.eu/eng/products/microwave-link.html}
\BIBentrySTDinterwordspacing

\bibitem{5523733}
P.~{Hasse}, D.~{Jaeger}, and J.~{Robert}, ``{DVB-C2} — a standard for
  improved robustness in cable networks,'' in \emph{Proceedings of the IEEE
  International Symposium on Consumer Electronics (ISCE 2010)}, 2010, pp. 1--6.

\bibitem{8808157}
B.~{Berscheid} and C.~{Howlett}, ``Full duplex {DOCSIS}: Opportunities and
  challenges,'' \emph{IEEE Communications Magazine}, vol.~57, no.~8, pp.
  28--33, Aug. 2019.

\bibitem{8954896}
Q.~Li, F.-K. Gong, P.-Y. Song, G.~Li, and S.-H. Zhai, ``{Beyond DVB-S2X:
  Faster-Than-Nyquist Signaling With Linear Precoding},'' \emph{IEEE
  Transactions on Broadcasting}, vol.~66, no.~3, pp. 620--629, Sep. 2020.

\bibitem{wen2021waveform}
S.~Wen, G.~Liu, C.~Liua, H.~Qu, M.~Tian, and Y.~Chen, ``Waveform design for
  high-order {QAM} faster-than-{Nyquist} transmission in the presence of phase
  noise,'' \emph{IEEE Transactions on Wireless Communications, {\normalfont
  Early Access},}, 2021 \color{black}.

\bibitem{song2020receiver}
P.~Song, F.~Gong, Q.~Li, G.~Li, and H.~Ding, ``Receiver design for
  faster-than-{Nyquist} signaling: Deep-learning-based architectures,''
  \emph{IEEE Access}, vol.~8, pp. 68\,866--68\,873, 2020 \color{black}.

\bibitem{souto2016mimo}
N.~Souto and R.~Dinis, ``{MIMO} detection and equalization for single-carrier
  systems using the alternating direction method of multipliers,'' \emph{IEEE
  Signal Processing Letters}, vol.~23, no.~12, pp. 1751--1755, Dec. 2016.

\bibitem{bedeer2017low}
E.~Bedeer, M.~H. Ahmed, and H.~Yanikomeroglu, ``Low-complexity detection of
  high-order {QAM} faster-than-{N}yquist signaling,'' \emph{IEEE Access},
  vol.~5, pp. 14\,579--14\,588, 2017.

\bibitem{bedeer2017very}
------, ``A very low complexity successive symbol-by-symbol sequence estimator
  for {faster-than-Nyquist} signaling,'' \emph{IEEE Access}, vol.~5, pp.
  7414--7422, 2017.

\bibitem{4373328}
L.~{Onural}, A.~{Gotchev}, H.~M. {Ozaktas}, and E.~{Stoykova}, ``A survey of
  signal processing problems and tools in holographic three-dimensional
  television,'' \emph{IEEE Transactions on Circuits and Systems for Video
  Technology}, vol.~17, no.~11, pp. 1631--1646, Oct. 2007.

\bibitem{liveris2003exploiting}
A.~D. Liveris and C.~N. Georghiades, ``Exploiting faster-than-{N}yquist
  signaling,'' \emph{IEEE Transactions on Communications}, vol.~51, no.~9, pp.
  1502--1511, Sep. 2003.

\bibitem{380028}
{Cheng-Kun Wang} and {Lin-Shan Lee}, ``Practically realizable digital
  transmission significantly below the {N}yquist bandwidth,'' \emph{IEEE
  Transactions on Communications}, vol.~43, no. 2/3/4, pp. 166--169,
  Feb./Mar/Apr. 1995.

\bibitem{4524864}
F.~{Rusek} and J.~B. {Anderson}, ``Non binary and precoded faster than
  {N}yquist signaling,'' \emph{IEEE Transactions on Communications}, vol.~56,
  no.~5, pp. 808--817, May 2008.

\bibitem{4801456}
F.~{Rusek}, ``On the existence of the {Mazo}-limit on {MIMO} channels,''
  \emph{IEEE Transactions on Wireless Communications}, vol.~8, no.~3, pp.
  1118--1121, Mar. 2009.

\bibitem{rusek2005two}
F.~Rusek and J.~B. Anderson, ``The two dimensional {M}azo limit,'' in
  \emph{Proceedings of the International Symposium on Information Theory
  (ISIT)}, 2005, pp. 970--974.

\bibitem{4939227}
F.~{Rusek} and J.~B. {Anderson}, ``Multistream faster than {N}yquist
  signaling,'' \emph{IEEE Transactions on Communications}, vol.~57, no.~5, pp.
  1329--1340, May 2009.

\bibitem{5288497}
A.~{Barbieri}, D.~{Fertonani}, and G.~{Colavolpe}, ``Time-frequency packing for
  linear modulations: spectral efficiency and practical detection schemes,''
  \emph{IEEE Transactions on Communications}, vol.~57, no.~10, pp. 2951--2959,
  Oct. 2009.

\bibitem{9120701}
A.~{Caglan}, A.~{Cicek}, E.~{Cavus}, E.~{Bedeer}, and H.~{Yanikomeroglu},
  ``Polar coded faster-than-{N}yquist ({FTN}) signaling with symbol-by-symbol
  detection,'' in \emph{Proceedings of the IEEE Wireless Communications and
  Networking Conference (WCNC)}, 2020, pp. 1--5.

\bibitem{6574905}
S.~{Sugiura}, ``Frequency-domain equalization of faster-than-{N}yquist
  signaling,'' \emph{IEEE Wireless Communications Letters}, vol.~2, no.~5, pp.
  555--558, Oct. 2013.

\bibitem{ibrahim2021novel}
A.~Ibrahim, E.~Bedeer, and H.~Yanikomeroglu, ``A novel low complexity
  faster-than-{Nyquist} signaling detector based on the primal-dual
  predictor-corrector interior point method,'' \emph{IEEE Communications
  Letters}, vol.~25, no.~7, pp. 2370--2374, Jul. 2021.

\bibitem{7510967}
T.~{Ishihara} and S.~{Sugiura}, ``Frequency-domain equalization aided iterative
  detection of faster-than-{N}yquist signaling with noise whitening,'' in
  \emph{Proceedings of the IEEE International Conference on Communications
  (ICC)}, 2016, pp. 1--6.

\bibitem{bedeer2019low}
E.~Bedeer, H.~Yanikomeroglu, and M.~H. Ahmed, ``Low-complexity detection of
  {M-ary PSK} faster-than-{N}yquist signaling,'' in \emph{Proceedings of the
  IEEE Wireless Communications and Networking Conference Workshop (WCNCW)},
  2019, pp. 1--5.

\bibitem{5205622}
J.~B. {Anderson}, A.~{Prlja}, and F.~{Rusek}, ``New reduced state space {BCJR}
  algorithms for the {ISI} channel,'' in \emph{Proceedings of the IEEE
  International Symposium on Information Theory}, 2009, pp. 889--893.

\bibitem{4595029}
A.~{Prlja}, J.~B. {Anderson}, and F.~{Rusek}, ``Receivers for
  faster-than-{N}yquist signaling with and without turbo equalization,'' in
  \emph{Proceedings of the IEEE International Symposium on Information Theory},
  2008, pp. 464--468.

\bibitem{6241379}
A.~{Prlja} and J.~B. {Anderson}, ``Reduced-complexity receivers for strongly
  narrowband intersymbol interference introduced by faster-than-{N}yquist
  signaling,'' \emph{IEEE Transactions on Communications}, vol.~60, no.~9, pp.
  2591--2601, Sep. 2012.

\bibitem{wen2020time}
S.~Wen, G.~Liu, Q.~Chen, H.~Qu, M.~Tian, J.~Guo, P.~Zhou, and D.~O. Wu,
  ``{Time-frequency compressed FTN signaling: A solution to spectrally
  efficient single-carrier system},'' \emph{IEEE Transactions on
  Communications}, vol.~68, no.~5, pp. 3125--3139, May 2020.

\bibitem{wen2019optimization}
S.~Wen, G.~Liu, Q.~Chen, H.~Qu, Y.~Wang, and P.~Zhou, ``{Optimization of
  precoded FTN signaling with MMSE-based turbo equalization},'' in
  \emph{Proceedings of the IEEE International Conference on Communications
  (ICC)}, May 2019, pp. 1--6.

\bibitem{jana2018dual}
M.~Jana, L.~Lampe, and J.~Mitra, ``{Dual-polarized faster-than-Nyquist
  transmission using higher order modulation schemes},'' \emph{IEEE
  Transactions on Communications}, vol.~66, no.~11, pp. 5332--5345, Nov. 2018.

\bibitem{jana2017pre}
M.~Jana, A.~Medra, L.~Lampe, and J.~Mitra, ``{Pre-equalized faster-than-Nyquist
  transmission},'' \emph{IEEE Transactions on Communications}, vol.~65, no.~10,
  pp. 4406--4418, Oct. 2017.

\bibitem{li2020code}
S.~Li, J.~Yuan, B.~Bai, and N.~Benvenuto, ``{Code-based channel shortening for
  faster-than-Nyquist signaling: Reduced-complexity detection and code
  design},'' \emph{IEEE Transactions on Communications}, vol.~68, no.~7, pp.
  3996--4011, Jul. 2020.

\bibitem{li2017reduced}
S.~Li, B.~Bai, J.~Zhou, P.~Chen, and Z.~Yu, ``{Reduced-complexity equalization
  for faster-than-Nyquist signaling: New methods based on Ungerboeck
  observation model},'' \emph{IEEE Transactions on Communications}, vol.~66,
  no.~3, pp. 1190--1204, Mar. 2018.

\bibitem{rusek2011optimal}
F.~Rusek and A.~Prlja, ``{Optimal channel shortening for MIMO and ISI
  channels},'' \emph{IEEE Transactions on Wireless Communications}, vol.~11,
  no.~2, pp. 810--818, Feb. 2012.

\bibitem{8798843}
A.~{Rashich} and S.~{Gorbunov}, ``{ZF} equalizer and trellis demodulator
  receiver for {SEFDM} in fading channels,'' in \emph{Proceedings of the
  International Conference on Telecommunications (ICT)}, 2019, pp. 300--303.

\bibitem{9006927}
W.~{Yuan}, N.~{Wu}, Q.~{Guo}, D.~W.~K. {Ng}, J.~{Yuan}, and L.~{Hanzo},
  ``Iterative joint channel estimation, user activity tracking, and data
  detection for {FTN-NOMA} systems supporting random access,'' \emph{IEEE
  Transactions on Communications}, vol.~68, no.~5, pp. 2963--2977, May 2020.

\bibitem{9185013}
Q.~{Li}, F.~{Gong}, P.~{Song}, G.~{Li}, and S.~{Zhai}, ``Joint channel
  estimation and precoding for faster-than-{N}yquist signaling,'' \emph{IEEE
  Transactions on Vehicular Technology}, vol.~69, no.~11, pp. 13\,139--13\,147,
  Sep. 2020.

\bibitem{8866862}
P.~{Song}, F.~{Gong}, and Q.~{Li}, ``Blind symbol packing ratio estimation for
  faster-than-{N}yquist signalling based on deep learning,'' \emph{Electronics
  Letters}, vol.~55, no.~21, pp. 1155--1157, Oct. 2019.

\bibitem{8301798}
J.~{Fan}, Y.~{Ren}, Y.~{Zhang}, and X.~{Luo}, ``Iterative carrier frequency
  offset estimation for faster-than-{N}yquist signaling,'' in \emph{Proceedings
  of the IEEE International Symposium on Wireless Personal Multimedia
  Communications (WPMC)}, 2017, pp. 150--153.

\bibitem{bedeer2017reduced}
E.~Bedeer, H.~Yanikomeroglu, and M.~H. Ahmed, ``Reduced complexity optimal
  detection of binary faster-than-{N}yquist signaling,'' in \emph{Proceedings
  of the IEEE International Conference on Communications (ICC)}, 2017, pp.
  1--6.

\bibitem{johnson2013statistical}
\BIBentryALTinterwordspacing
D.~H. Johnson, ``Statistical signal processing,'' \emph{Lecture Notes}, 2013,
  last accessed on 10-11-2020. [Online]. Available:
  \url{http://cnx.org/content/col11382/1.1/. Lecture notes}
\BIBentrySTDinterwordspacing

\bibitem{birgin2014practical}
E.~G. Birgin and J.~M. Mart{\'\i}nez, \emph{Practical Augmented Lagrangian
  Methods for Constrained Optimization}.\hskip 1em plus 0.5em minus 0.4em\relax
  SIAM, 2014.

\bibitem{bertsekas2014constrained}
D.~P. Bertsekas, \emph{Constrained Optimization and Lagrange Multiplier
  Methods}.\hskip 1em plus 0.5em minus 0.4em\relax Academic Press, 2014.

\bibitem{takapoui2020simple}
R.~Takapoui, N.~Moehle, S.~Boyd, and A.~Bemporad, ``A simple effective
  heuristic for embedded mixed-integer quadratic programming,''
  \emph{International Journal of Control}, vol.~93, no.~1, pp. 2--12, Apr.
  2017.

\bibitem{boyd2011distributed}
S.~Boyd, N.~Parikh, E.~Chu, B.~Peleato, and J.~Eckstein, ``Distributed
  optimization and statistical learning via the alternating direction method of
  multipliers,'' \emph{Foundations and Trends{\textregistered} in Machine
  Learning}, vol.~3, no.~1, pp. 1--122, 2011.

\bibitem{luo1993convergence}
Z.-Q. Luo and P.~Tseng, ``On the convergence rate of dual ascent methods for
  linearly constrained convex minimization,'' \emph{Mathematics of Operations
  Research}, vol.~18, no.~4, pp. 846--867, Nov. 1993.

\bibitem{1523702}
M.~Sikora and D.~Costello, ``{A new SISO algorithm with application to turbo
  equalization},'' in \emph{Proceedings of IEEE International Symposium on
  Information Theory (ISIT)}, 2005, pp. 2031--2035.

\bibitem{1194444}
B.~M. {Hochwald} and S.~{ten Brink}, ``Achieving near-capacity on a
  multiple-antenna channel,'' \emph{IEEE Transactions on Communications},
  vol.~51, no.~3, pp. 389--399, Mar. 2003.

\bibitem{7968313}
T.~Ishihara and S.~Sugiura, ``{Iterative Frequency-Domain Joint Channel
  Estimation and Data Detection of Faster-Than-Nyquist Signaling},'' \emph{IEEE
  Transactions on Wireless Communications}, vol.~16, no.~9, pp. 6221--6231,
  Sep. 2017.

\bibitem{kulhandjian2019low}
M.~Kulhandjian, E.~Bedeer, H.~Kulhandjian, C.~D’Amours, and H.~Yanikomeroglu,
  ``Low-complexity detection for faster-than-{N}yquist signaling based on
  probabilistic data association,'' \emph{IEEE Communications Letters},
  vol.~24, no.~4, pp. 762--766, Apr. 2020.

\end{thebibliography}
\end{document}